\documentclass[trackchanges]{aastex701}
\usepackage{amsmath}
\begin{document}

\title{Backmapping of the High- and Low-latitude Solar Wind under Multiple Heliospheric and Coronal Magnetic Field Configurations}

\correspondingauthor{Liping Yang, Xueshang Feng}
\email{lpyang@swl.ac.cn, fengx@spaceweather.ac.cn}

\author{Xinyi Ma}
\affiliation{State Key Laboratory of Solar Activity and Space Weather, National Space Science Center, Chinese Academy of Sciences, Beijing 100190, People's Republic of China}
\affiliation{College of Earth and Planetary Sciences, University of Chinese Academy of Sciences, Beijing 100049, People's Republic of China}
\email{ }

\author[0000-0003-4716-2958]{Liping Yang}
\affiliation{State Key Laboratory of Solar Activity and Space Weather, National Space Science Center, Chinese Academy of Sciences, Beijing 100190, People's Republic of China}
\email{ }

\author[0000-0001-8605-2159]{Xueshang Feng}
\affiliation{State Key Laboratory of Solar Activity and Space Weather, National Space Science Center, Chinese Academy of Sciences, Beijing 100190, People's Republic of China}
\affiliation{Shenzhen Key Laboratory of Numerical Prediction for Space Storm, School of Aerospace, Harbin Institute of Technology, Shenzhen 518055, People's Republic of China}
\email{ }

\author[0000-0002-1369-1758]{Hui Tian}
\affiliation{School of Earth and Space Sciences, Peking University, 100871 Beijing, People's Republic of China}
\affiliation{State Key Laboratory of Solar Activity and Space Weather, National Space Science Center, Chinese Academy of Sciences, Beijing 100190, People's Republic of China}
\email{ }

\author[0000-0003-0424-9228]{Honghong Wu}
\affiliation{School of Electronic Information, Wuhan University, 430072 Wuhan,  People's Republic of China}
\email{ }

\author[0000-0002-4935-6679]{Fang Shen}
\affiliation{State Key Laboratory of Solar Activity and Space Weather, National Space Science Center, Chinese Academy of Sciences, Beijing 100190, People's Republic of China}
\affiliation{College of Earth and Planetary Sciences, University of Chinese Academy of Sciences, Beijing 100049, People's Republic of China}
\email{ }

\author{Wangning Zhang}
\affiliation{State Key Laboratory of Solar Activity and Space Weather, National Space Science Center, Chinese Academy of Sciences, Beijing 100190, People's Republic of China}
\affiliation{School of Geophysics and Information Technology, China University of Geosciences (Beijing), China}
\email{ }

\author{Mengxuan Ma}
\affiliation{State Key Laboratory of Solar Activity and Space Weather, National Space Science Center, Chinese Academy of Sciences, Beijing 100190, People's Republic of China}
\affiliation{College of Earth and Planetary Sciences, University of Chinese Academy of Sciences, Beijing 100049, People's Republic of China}
\email{ }

\author{Xiao Zhang}
\affiliation{State Key Laboratory of Solar Activity and Space Weather, National Space Science Center, Chinese Academy of Sciences, Beijing 100190, People's Republic of China}
\affiliation{College of Earth and Planetary Sciences, University of Chinese Academy of Sciences, Beijing 100049, People's Republic of China}
\email{ }

\author{Ziwei Wang}
\affiliation{Shenzhen Key Laboratory of Numerical Prediction for Space Storm, School of Aerospace, Harbin Institute of Technology, Shenzhen 518055, People's Republic of China}
\email{ }

\begin{abstract}

Solar wind backmapping is a critical technique for analyzing the origin of the solar wind and space weather events by correlating in situ measurements with solar remote-sensing observations. This technique typically traces magnetic field lines using a heliospheric magnetic field (HMF) model coupled with a coronal magnetic field (CMF). However, the impact of different HMF and CMF configurations on backmapping uncentainty-particularly regarding high-latitude solar wind-remains inadequately quantified.
This study comprehensively evaluates solar wind backmapping by combining two HMF models (Parker spiral, Fisk-type) with three CMF models (Potential Field Source Surface (PFSS), Potential Field Current Sheet (PFCS), Current Sheet Source Surface (CSSS)). Our analysis primarily uses in situ measurements from Ulysses and remote-sensing data from STEREO-A. Key findings are that: (1) while both Fisk and Parker HMF models show comparable consistency with measured magnetic field strength and polarity, they produce certain longitudinal displacements in their back-mapped footpoints on the source surface (2.5$R_{\odot}$); (2) For CMF models (PFSS, PFCS, CSSS), predicted photospheric footpoints exhibit minor variations for high/mid-latitude solar wind but some divergences for ecliptic/low-latitude wind; (3) All three CMF models link high/mid-latitude wind to active regions or coronal holes, yet associate a fraction of ecliptic/low-latitude wind with quiet-Sun regions; (4) Ecliptic/low-latitude sources show significantly stronger dependence on the PFSS source surface height compared to high-latitude wind.
These results demonstrate that simpler models (PFSS + Parker) appear reasonably adequate for polar coronal hole wind studies, while low-latitude/ecliptic solar wind exhibits the heightened sensitivity to model choices.

\end{abstract}


%\keywords{\uat{Galaxies}{573} --- \uat{Cosmology}{343} --- \uat{High Energy astrophysics}{739} --- \uat{Interstellar medium}{847} --- \uat{Stellar astronomy}{1583} --- \uat{Solar physics}{1476}}


\section{Introduction} \label{sec:intro}
Solar wind backmapping provides critical insights into solar-terrestrial physics, such as determining the origin of the solar wind, enabling space weather forecasting \citep{2013Owens,2025Cranmer}.
Magnetic field configurations provide the essential topological linkage between dynamic solar phenomena and their manifestations in interplanetary space. By inversely tracing solar wind parcels measured in situ by spacecraft to their source locations on the solar surface, researchers can elucidate critical processes governing solar wind generation, acceleration, and heating mechanisms \citep{1973Nolte, Levine1977, Pizzo1981, Neugebauer1998, Linker1999, Yang2012, 2017Peleikis, Rouillard2020, Parenti2021, Bale2023, Kumar2023, Rivera2024, Ervin2024, Hou2024a, Hou2024b, Bizien2025}.

Multiple methodologies facilitate magnetic connectivity across heliospheric domains.
Some techniques extrapolate solar wind parameters near the Sun to a chosen heliospheric position, while others relate in situ measurements far from the Sun back to the solar atmosphere \citep{2011Riley}.
Prominent methods include ballistic extrapolation \citep{2013Balogh}, ad hoc kinetic mapping \citep{2000Arge}, 1-D upwinding propagation \citep{2011Riley}, global magnetohydrodynamics (MHD) simulations \citep{Posner2001}, and so on. Among these, ballistic solar wind backmapping assumes the constant-speed radial flow of the solar wind and remains the most extensively utilized operational approach \citep{2013Balogh,2013Owens,2017Peleikis,2021Macneil}.
This technique commonly combines the Potential Field Source Surface (PFSS) coronal magnetic field (CMF) model with the Parker spiral heliospheric magnetic field (HMF) model to trace in situ solar wind measurements inward to the solar disk.

The PFSS model assumes that the coronal magnetic field is potential and extrapolates the coronal magnetic structure from photospheric magnetic field observations. It further assumes the coronal field becomes radially open beyond a source surface, typically set at 2.5 solar radii \citep{1969Altschuler}. The PFSS model is widely used to analyze coronal magnetic topology, study coronal large-scale structures, predict solar wind properties, and other applications \citep{2000Arge,2023Lin}. However, it ignores the effect of  electric currents \citep{1969Schatten}. To address this limitation, alternative CMF models like the Potential Field Current Sheet (PFCS) and Current Sheet Source Surface (CSSS) models have been developed. The PFCS model incorporates current sheets at magnetic polarity reversal sites, and can improves the directions and variations of magnetic fields to match remote sensing and in situ observations \citep{1971Schatten, 2024Shi}.
The CSSS model accounts for horizontal coronal currents, producing notably smoother and more continuous open magnetic field lines than the PFSS model \citep{1995Zhao,2019Koskela}.
These alternative CMF models can alter open-flux geometry, thereby modifying solar wind source mapping.

Beyond the source surface, solar wind backmapping relies on HMF models. The Parker spiral model is a widely used HMF approach, assuming magnetic fields are rooted at the Sun and carried outward by the steady, radial flow of the solar wind into interplanetary space \citep{1958Parker}.
This model describes how solar rotation twists magnetic fields into Archimedean spirals, which are confined to cones of constant heliographic latitude.
When tracing solar wind measurements through the Parker spiral to the boundary of a CMF model, the ballistic approximation treats the solar wind with a constant speed, radial, plasma flow. Some studies enhance this approximation by incorporating solar wind acceleration and non-radial flow components \citep{2021Macneil,2024Dakeyo,2024ShiXinzheng}. These analyses reveal that solar wind acceleration and corotation effects counterbalance each other, which ultimately leads to a good performance of ballistic backmapping.

To explain recurrent energetic particle events observed at high latitudes by the Ulysses spacecraft, \citet{1996Fisk} proposed an alternative to the Parker spiral HMF model that permits large excursions of magnetic fields in latitude. Based on a tilted  magnetic dipole, the Fisk HMF model includes the interplay between footpoint differential rotation and magnetic field superradial expansion \citep{1997Zurbuchen,1999Fisk,2003Schwadron,2008Burger,2020Steyn}.
To our knowledge, few studies have implemented the Fisk model in the solar wind backmapping, with \citet{2024Steyn} being the only application to traces field lines from L1 to the photosphere. Their results demonstrate exclusive mapping of the Fisk model to polar regions, contrasting with the PFSS model's mapping to equatorial and mid-latitude photospheric locations. By anchoring HMF footpoints in the photosphere and modeling their nonradial expansion through rigidly rotating coronal holes (predominantly polar), the Fisk model is particularly well-suited for tracing the fast solar wind back to its solar polar sources.

Despite widespread implementation, backmapping methodologies retain critical limitations.
Ballistic backmapping is predominantly employed to relate ecliptic solar wind measurements with their coronal sources like mid-latitude coronal holes and active regions. However,it remains largely unexplored to connect back the high-latitude solar wind, which is typically associated with polar coronal holes. This knowledge gap is increasingly consequential given Solar Orbiter's out-of-ecliptic orbits \citep{2020Muller} and China's planned solar polar-orbit observatory mission \citep{2025Deng}. Furthermore, only the PFSS model and Parker models are widely implemented in backmapping studies. A rigorous evaluation of how different CMF and HMF models influence the identification of solar wind source region is warranted.

In this study, we employ the PFSS, PFCS, and CSSS CMF models alongside the Parker and Fisk HMF models to trace back solar wind measurements from Ulysses and Wind spacecraft to their solar origins. The resulting source locations are compared with Extreme Ultraviole Imagers (EUVI) observations from the Solar Terrestrial Relations Observatory Ahead spacecraft (STEREO-A). We quantify angular deviations between backmapping methods using different CMF/HMF model combinations, with particular focus on the fast solar wind originating from the Sun's polar regions. The paper is structured as follows: Section \ref{sec:data} details the observational data and the modeling frameworks. Section \ref{sec:results} presents and analyzes backmapping results across the CMF (PFSS, PFCS, CSSS) and HMF (Parker, Fisk) model configurations as well as different source surface heights of the PFSS model. Section \ref{sec:conclu} provides concluding remarks and discussion.

\section{Data and Models} \label{sec:data}

\subsection{In Situ Measurements} \label{subsec:obser}

As the first deep space probe to achieve a high-inclination orbit around the Sun, Ulysses completed the south pole crossing in early 2007 and arrived at the north pole at the end of the same year, conducting in situ measurements of the solar wind at latitudes $>60^{\circ}$. The magnetic field and plasma parameters were obtained by the Vector Helium Magnetometer (VHM) and Solar Wind Observations Over the Poles of the Sun (SWOOPS) \citep{1992Balogh}, respectively. In this work, we mainly use hour-averaged magnetic field and plasma data measured by the Ulysses spacecraft during its 2007 polar orbit, covering a heliocentric distance range of 1.4 to 2.3 AU and a latitude range from $80^{\circ}S$ to $80^{\circ}N$.

The Wind spacecraft locates at the L1 Lagrange point of the solar-terrestrial system, and continuously monitors the solar wind environment at upside of the earth. It records the temporal evolution of magnetic field, flow velocity, density and other parameters of the solar wind through the Magnetic Field Investigation (MFI) and the Solar Wind Experiment (SWE) \citep{2021Wilson}. In this work, we use hour-averaged magnetic field and plasma data measured by Wind spacecraft from September 10 to October 8, 2007 \citep{King_Papitashvili_2020}. 

\subsection{Remote Sensing Observations}

Although promissing advances have been made recently in the measurements of the global coronal magnetic field \citep{2020Yanga,2020Yangb,2024Yangc}, measurements of the 3D global magnetic field structure in the entire solar atmosphere are still not possible and we need to rely on the technique of magnetic field extrapolation.
We utilize daily synoptic photospheric magnetograms generated by the Air Force Data Assimilative Photospheric Flux Transport (ADAPT) model \citep{Arge2010} as the boundary condition for solar coronal magnetic field extrapolation. These ADAPT maps assimilate raw full-disk magnetograms obtained by the Global Oscillation Network Group (GONG). The GONG project maintains continuous solar monitoring through a global six-station network, synthesizing daily synoptic magnetograms (spatial resolution: 180 latitude $\times$ 360 longitude points in $\sin\theta$-$\phi$ coordinates). This multi-station strategy overcomes single-site limitations imposed by Earth's rotation and weather disruptions, enabling near-continuous photospheric magnetic coverage. Through ADAPT's flux transport physics, the assimilated maps provide dynamically consistent photospheric evolution of the magnetic field. In this work, we employs ADAPT maps covering seven CRs in 2007 and CR 2066 in 2008 to extrapolate coronal magnetic fields.

Additionally employed are observations of the lower corona obtained from the EUVI of the Sun Earth Connection Coronal and Heliospheric Investigation (SECCHI) suite onboard the STEREO-A spacecraft \citep{2008Howard}. EUVI observations can reveal coronal hole regions, active regions, quiet-Sun regions, and are thus applied to evaluate the accuracies of the mapped source regions of the solar wind \citep{Yang2012,2013Owens}. In this work, we utilize synoptic maps derived from STEREO-A 195 $\mathring{A}$ full-disk EUVI observations. These maps were assembled from these full-disk extreme ultraviolet (EUV) image acquired throughout 2007, during which STEREO-A was positioned ahead of Earth in its heliocentric orbit.

\subsection{Coronal Magnetic Field Models} \label{subsec:CMF}

\subsubsection{PFSS CMF Model}
The PFSS model extrapolates the solar coronal magnetic field by assuming a current-free approximation in the region between the photosphere and the source surface \citep{1992Wang, Li2021}. This assumption leads to the condition:
\begin{equation}
\nabla \times \boldsymbol{B}=0,
\end{equation}
consequently, the magnetic field $\boldsymbol{B}$ can be represented as the negative gradient of a scalar magnetic potential:
\begin{equation}
\boldsymbol{B}=-\nabla \Psi_B.
\end{equation}
Combining with the divergence-free condition of magnetic field:
\begin{equation}
\nabla \cdot \boldsymbol{B} =0,
\end{equation}
yields Laplace's equation for the potential:
\begin{equation}
\nabla^2 \Psi_B =0.
\end{equation}
The general solution to Laplace's equation in spherical coordinates ($r$, $\theta$, $\phi$) takes the form of a spherical harmonic expansion \citep{1962Jackson}:
\begin{equation}
\Psi_B(r, \theta, \phi) = \sum_{l=0}^{\infty} \sum_{m=-l}^{l} \left[ f_l^m r^l + g_l^m r^{-(l+1)} \right] P_l^m(\cos\theta) e^{imf},
\end{equation}
where $P_l^m$ are the Legendre polynomials, and the coefficients $f_l^m$, $g_l^m$ are determined by boundary conditions.
The model employs two physically motivated boundary conditions:\\
1. Photospheric Boundary ($r = R_\odot$): \\
The radial field component $B_r(R_\odot, \theta, \phi)$ is specified using synoptic magnetogram data, providing the primary input for the solution.\\
2. Source Surface Boundary ($r = R_\mathrm{{SS}}$): \\
On the source surface (conventionally $R_\mathrm{{SS}} = 2.5 R_\odot$), the magnetic field is constrained to become purely radial:
\begin{equation}
B_\theta(R_\mathrm{{SS}}, \theta, \phi) = B_\phi(R_\mathrm{{SS}}, \theta, \phi) = 0.
\end{equation}

\subsubsection{PFCS CMF model}
The PFCS model represents a refinement of the PFSS model by accounting for the horizontal current effect in the corona,
and thus reproduce the latitudinally independent radial magnetic fields
measured by Ulysses \citep{Smith1995}. Within the inner region ($r < R_\mathrm{{SS}}$), the PFCS model assumes that the CMFs are current free, which means that magnetic fields derive from a scalar potential. When transiting to the outer region ($R_{\mathrm{SS}} < r < R_{\mathrm{SCS}}$, with $R_{\mathrm{SCS}}=13 R_{\odot}$ being the outer boundary of the PFCS model), the potential extrapolation is also used, with the radial magnetic field $B_r$ on the source surface serving as the lower boundary condition.
To induce the horizontal current, a polarity reversal method is used in the outer region. Specifically,
If $B_r>0$ on the source surface, $B_\theta$ and $B_\phi$ are unchanged.
If $B_r<0$ on the source surface, $B_r=-B_r$, $B_\theta=-B_\theta$ and $B_\phi=-B_\phi$.
The thin horizontal current sheet condition is introduced in the opposite field region \citep{1971Schatten}:
\begin{equation}
\Psi_B(r, \theta, \phi) = R \sum_{l=0}^{\infty} \sum_{m=0}^{l} (\frac{R}{r})^{n+1} \left[ g^m_l \cos(m\phi) + h^m_l sin(m\phi) \right] P_l^m(\theta),
\end{equation}
where the coefficients $g^m_l$ and $h^m_l$ are determined by boundary conditions.
The original orientations of the magnetic fields are restored after the magnetic fields are obtained in the outer region.

\subsubsection{CSSS CMF Model}
Unlike the PFSS and PFCS models, the CSSS model includes two spherical surfaces: the cusp surface (located at $r=R_{\mathrm{CS}}$, approximately the height of the cusp point of the coronal helmet streamers) and the source surface. From the solar surface to the cusp surface which is called the inner region, the scalar magnetic potential $\Psi$ is expressed as \citep{1994Zhao,1995Zhao,2024Shi}:
\begin{equation}
\Psi_B(r, \theta, \phi) = \sum_{l=1}^{N_{\odot}} \sum_{m=0}^{l} R^{\odot}_l(r) P^m_l(\cos\theta) \left[ g^{\odot}_{\mathrm{ml}} \cos(m\phi) + h^{\odot}_{\mathrm{ml}}  sin(m\phi) \right],
\end{equation}
\begin{equation}
R^{\odot}_l(r)=\frac{R_{\odot}(1+a)^n}{(n+1)(r+a)^{n+1}},
\end{equation}
where $N_{\odot}$ is the maximum principal index, the coefficients $g^{\odot}_{ml}$ and $h^{\odot}_{ml}$ are determined by the photospheric synoptic maps.

From the cusp surface to the source surface which is called the middle region, the magnetic fields gradually orient toward the
radial direction. The scalar magnetic potential in this region satisfies \citep{1994Zhao,1995Zhao,2024Shi}:
\begin{equation}
\Psi_B(r, \theta, \phi) = \sum_{l=0}^{N_c} \sum_{m=0}^{l} R^c_l(r) P^m_l(\cos\theta) \left[ g^c_{\mathrm{ml}} \cos(m\phi) + h^c_{\mathrm{ml}} sin(m\phi) \right],
\end{equation}
\begin{equation}
R^c_{\mathrm{ml}}(r)=R_{\odot}\left[ \frac{n+1}{R^2_{\mathrm{CS}}(R_{\mathrm{CS}}+a)^n} + \frac{n(R_{\mathrm{CS}}+a)^{n+1}}{R^2_{\mathrm{CS}}(R_{\mathrm{SS}}+a)^{2n+1}}    \right]^{-1} \\
\times \left[\frac{1}{(r+a)^{n+1}}-\frac{(r+a)^n}{(R_{\mathrm{SS}}+a)^{2n+1}}\right],
\end{equation}
where $N_c$ is the maximum principal index, the coefficients $g^c_{\mathrm{ml}}$ and $h^c_{\mathrm{ml}}$ are determined by boundary conditions on the cusp surface.

Beyond the source surface which is called the outer region, the magnetic fields are completely open and follow Parker spiral lines. According to \citet{2024Shi}, we set the cusp surface height $R_{\mathrm{CS}}=2.5R_{\odot}$, the source surface height $R_{\mathrm{SS}}=13R_{\odot}$ and the current parameter $a=1.0R_{\odot}$ in the CSSS model.

\subsubsection{Fisk CMF Model}

Fisk model assumes magnetic fields expand from polar coronal holes to the source surface symmetrically about the magnetic axis. \citet{1996Fisk} proposed the transformation between the photosphere's footpoints and the source surface's footpoints. First, the footpoints on the source surface have heliomagnetic coordinate given by
\begin{equation}
\begin{aligned}
& \theta_{\mathrm{ss}}^{\mathrm{hm}}=\cos^{-1}(\sin(\theta_{\mathrm{ss}}^{\mathrm{hg}})\cos(\phi_{\mathrm{ss}}^{\mathrm{hg}})\sin(\alpha)+\cos(\theta_{\mathrm{ss}}^{\mathrm{hg}})\cos(\alpha)), \\
& \phi_{\mathrm{ss}}^{\mathrm{hm}}=\cos^{-1}(\frac{\sin(\theta_{\mathrm{ss}}^{\mathrm{hg}})\cos(\phi_{\mathrm{ss}}^{\mathrm{hg}})\cos(\alpha)-\cos(\theta_{\mathrm{ss}}^{\mathrm{hg}})\sin(\alpha)}{\sin(\theta_{\mathrm{ss}}^{\mathrm{hm}})}).
\end{aligned}
\end{equation}
Then, based on the Divergence Theorem, the footpoints on the photosphere tracing from the source surface are obtained by
\begin{equation}
\begin{aligned}
& \theta_{\mathrm{ph}}^{\mathrm{hm}} = \sin^{-1} \left(
  \sqrt{
    \frac{
      (1 - \cos(\theta_{\mathrm{ss}}^{\mathrm{hm}})) \sin^2(\theta_{\mathrm{ph}}^{\mathrm{mm}})
    }{
      (1 - \cos(\theta_{\mathrm{ss}}^{\mathrm{mm}}))
    }
    \vphantom{\dfrac{1}{1}}
  }
\right), \\
& \phi_{\mathrm{ph}}^{\mathrm{hm}}=\phi_{\mathrm{ss}}^{\mathrm{hm}}.
\end{aligned}
\end{equation}
Finally, we can obtain the footpoints in the heliographic coordinate system by
\begin{equation}
\begin{aligned}
& \theta_{\mathrm{ph}}^{\mathrm{hg}}=\cos^{-1}(\cos(\theta_{\mathrm{ph}}^{\mathrm{hm}})\cos(\alpha)-\sin(\theta_{\mathrm{ph}}^{\mathrm{hm}})\cos(\phi_{\mathrm{ph}}^{\mathrm{hm}})\sin(\alpha)), \\
& \phi_{\mathrm{ph}}^{\mathrm{hg}}=\cos^{-1}(\frac{\sin(\theta_{\mathrm{ph}}^{\mathrm{hm}})\cos(\phi_{\mathrm{ph}}^{\mathrm{hm}})\cos(\alpha)+\cos(\theta_{\mathrm{ph}}^{\mathrm{hm}})\sin(\alpha)}{\sin(\theta_{\mathrm{ph}}^{\mathrm{hm}})}),
\end{aligned}
\end{equation}
where the subscripts ${\mathrm{ss}}$, ${\mathrm{ph}}$, ${\mathrm{hg}}$ and ${\mathrm{hm}}$ represents the source surface, photosphere, heliographic coordinates, and heliomagnetic coordinates, respectively. $\theta_{\mathrm{ph}}^{\mathrm{mm}}$ and $\theta_{\mathrm{ss}}^{\mathrm{mm}}$ are chosen to be $24^{\circ}$ and $70^{\circ}$ \citep{2024Steyn}. It is worth noting that the Fisk coronal model only provides the correspondence between the source surface and the footpoints of the photosphere, and does not provide a specific expression for the coronal magnetic field.

\subsection{Heliospheric Magnetic Field Models} \label{subsec:HMF}
\subsubsection{Parker HMF Model}
Parker HMF model describes the spiral structure imposed on the solar wind by the Sun's rotation \citet{1958Parker}. The field components in the Heliocentric Carrington Coordinate System are expressed as:
\begin{equation} \label{Parker}
\begin{aligned}
& B_r=B_0(\frac{r_0}{r})^2 \\
& B_\theta=0 \\
& B_\phi=- \frac{B_r r}{V} \Omega \sin\theta ,
\end{aligned}
\end{equation}
where $B_0$ represents the radial field strength at the reference distance $r_0$, $V$ is the solar wind speed, $\Omega$ is the angular rotation velocity of the Sun; $r$, $\theta$ and $\phi$ are the heliospheric distance, colatitude, and longitude, respectively. The radial component $B_r$ decays with the inverse square of heliocentric distance, reflecting magnetic flux conservation in a radially expanding solar wind of constant speed $V$. The azimuthal component $B_\phi$ emerges from the $\Omega$, which causes the magnetic field lines to take an Archimedean spiral configuration.

\subsubsection{Fisk HMF Model}
Fisk model extends the classical Parker spiral by accounting for the sun's differential rotation on the photosphere and non-radial excursions. The HMF field components in the Heliocentric Carrington Coordinate System are detailed by \citet{1997Zurbuchen} as follows:
\begin{equation} \label{Fisk}
\begin{aligned}
& B_r=B_0(\frac{r_0}{r})^2 \\
& B_\theta=\frac{B_0 {r_0}^2}{V r} \omega \sin\beta \sin(\phi+ \frac{\Omega r}{V} - \phi_0 )\\
& B_\phi=\frac{B_0 {r_0}^2}{V r} [\omega(\cos\beta \sin\theta + \sin\beta \cos\theta \cos(\phi+ \frac{\Omega r}{V} - \phi_0 ) ) -\Omega \sin\theta ],
\end{aligned}
\end{equation}
where $B_0$ represents the radial field strength at the reference distance $r_0$, $V$ is the solar wind speed, $\Omega$ is the angular rotation velocity of the sun. The key parameters governing the magnetic field topology are: $\beta$ is the polar angle between a magnetic field line p which originates from the rotational pole, after crossing the source surface and the axis of the rotation, $\omega$ is the differential rotation rate of the footpoints on the photosphere, $\alpha$ is the tilt angle of the solar magnetic axis relative to the rotation axis, $\phi_0$ is the plane formed by the rotation axis and the magnetic axis. As previous works \citep{2002Forsyth}, we adapt the specific parameter values  $\beta=30^\circ$, $\alpha=15^\circ$, $\omega=\frac{\Omega}{4}$ and $\phi_0=100^\circ$.

\section{Results} \label{sec:results}
We select three Carrington Rotations (CRs) to present detailed backmapping analysis:
CR 2060 (August 14 - September 10, 2007), CR 2061 (September 10 - October 8, 2007), and CR 2064 (December 1 - December 29, 2007).
During these periods, Ulysses was positioned at $\sim1.41$, $\sim1.46$, $\sim1.88$ AU and latitudes of $3.3^{\circ}N$ to $23.1^{\circ}N$, $23.1^{\circ}N$ to $42.1^{\circ}N$, $70.1^{\circ}N$ to $78.3^{\circ}N$ respectively.
Our solar wind backmapping methodology proceeds as follows:
First, we quantify agreement in strength, polarity, and configuration between Parker/Fisk-predicted and Ulysses/Wind-observed HMFs.
Then, we trace interplanetary solar wind back to CMF model boundaries using HMF models, evaluating spatial separations between Parker- and Fisk-mapped points.
Finally, using the CMF models (PFSS, PFCS, CSSS, Fisk), we identify solar wind source locations, which are validated through EUVI observations. We also compare results across the CMF models.

\subsection{Mid-latitude and Ecliptic Backmapping}

\begin{figure*}[htbp]
  \centering
  \includegraphics[width=1\textwidth]{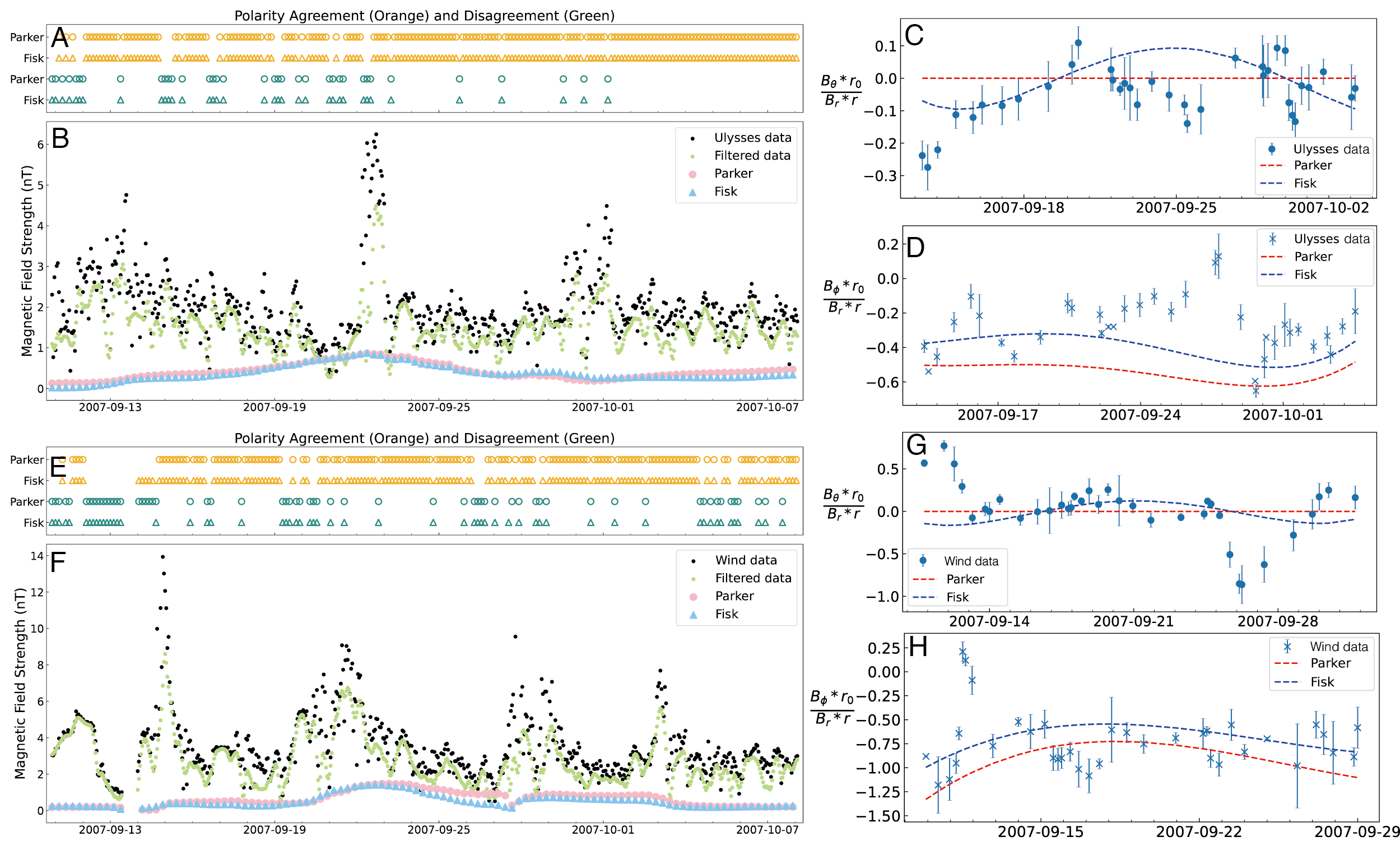}
  \caption{Predicted HMFs by the Parker and Fisk models with in situ measurements obtained from the Ulysses (A, B, C, D) and Wind (E, F, G, H).
  Panels A and E: the agreement (orange) and disagreement (green) between measured and predicted HMF polarity.
  Panels B and F: black dots and light green dots denote one-hour-cadence and filtered measured magnetic field strength, respectively; pink dots and lightskyblue dots shows the predicted magnetic field strength by the Parker and Fisk models, respectively.
  Panels C and G: the observed (circular markers) and predicted values (red dotted line for Parker and blue dotted line for Fisk) of $B_\theta*r_0 / B_r*r$ (where $r_0= 1$ AU and r is the heliocentric distance), representing the meridional magnetic field component $B_\theta$ divided by the radial magnetic field component $B_r$.
  Panels D and H: the same as Panels C and G but for azimuthal field component ($B_\phi$).}
  \label{fig:1}
\end{figure*}

To evaluate the fidelity of Parker and Fisk heliospheric HMF models, we compare predicted and observed HMF polarity, strength, and configuration in Figure \ref{fig:1}. Panels (A) and (E) quantify polarity agreement (orange) and disagreement (green) between model predictions and spacecraft observations, determined through sign comparisons of the radial magnetic field component. For Ulysses' mid-latitude measurements, both models achieve identical polarity match rates of $83.61\%$ with observations. In the ecliptic plane, Wind observations show $70.67\%$ agreement with Parker predictions and $73.01\%$ with Fisk predictions.

We show in Figure \ref{fig:1}(B) and (F) one-hour-cadence (black dots) and filtered (light green dots) magnetic field strength measured by the Ulysses (B) and Wind (F) spacecraft as well as the magnetic field strength obtained from the Parker (pink dots) and Fisk models (lightskyblue dots). Raw magnetic field maps were preprocessed using Gaussian filtering to attenuate high-frequency noise while preserving field gradients. This operation was implemented via $scipy.ndimage.gaussian\_filter$ in Python, generating smoothed field strength distributions for subsequent analysis.
The unsigned residual metric $\chi$ \citep{2024Shi} is used to quantify the consistency of the magnetic field strength between measurements and models, and is defined by the expression:
\begin{equation}
\label{eq:chi}
\chi = \frac{1}{N} \sum_{i=1}^{N} \sqrt{(|M_i|-|O_i|)^2},
\end{equation}
where $N$ is the number of measured data, $M$ denotes the magnetic field strength of the predicted model, and $O$ represents the filtered observational data. Smaller residual values indicate better agreement with observations.

For the mid-latitude measurements by Ulysses, the residuals for the Parker and Fisk models are 1.09 nT and 1.12 nT, respectively. In the ecliptic plane, the residuals for the Parker and Fisk models are 2.09 nT and 2.19 nT, respectively. The Parker and Fisk models exhibits better consistency with observations at the mid-latitude than at the ecliptic plane, though its predictions systematically underestimate the measured values.

Figure \ref{fig:1}(C) and (G) display the observed (circular markers) and predicted values (red dotted line for Parker and blue dotted line for Fisk) of $B_\theta*r_0 / B_r*r$ (where $r_0= 1$ AU and r is the heliocentric distance), representing the meridional magnetic field component $B_\theta$ divided by the radial magnetic field component $B_r$.  Figure \ref{fig:1}(D) and (H) show the same as Figure \ref{fig:1}(C) and (G) but for the azimuthal magnetic field component $B_\phi$.
$B_\theta*r_0 / B_r*r$ and $B_\phi*r_0 / B_r*r$ are calculated using the method described by \citet{1997Zurbuchen}, following these steps:\\
1. High-frequency variations are suppressed using a Gaussian filter.\\
2. The Carrington longitudes on the source surface are computed from the in situ solar wind speed.\\
3. Data are binned into $10^\circ$ Carrington longitude intervals.\\
4. A 3-point running average is applied for smoothing.

Error bars represent the standard error of the mean within each Carrington longitude bin.

While neither model fully captures the observed complexity of ecliptic plane to mid-latitude field configurations \citep{Park2025}, the Fisk model demonstrates superior alignment with measurements in both amplitude and phase compared to the Parker model. This improvement is particularly evident in its representation of the meridional field component.

\begin{figure*}[htbp]
  \centering
  \includegraphics[width=1\textwidth]{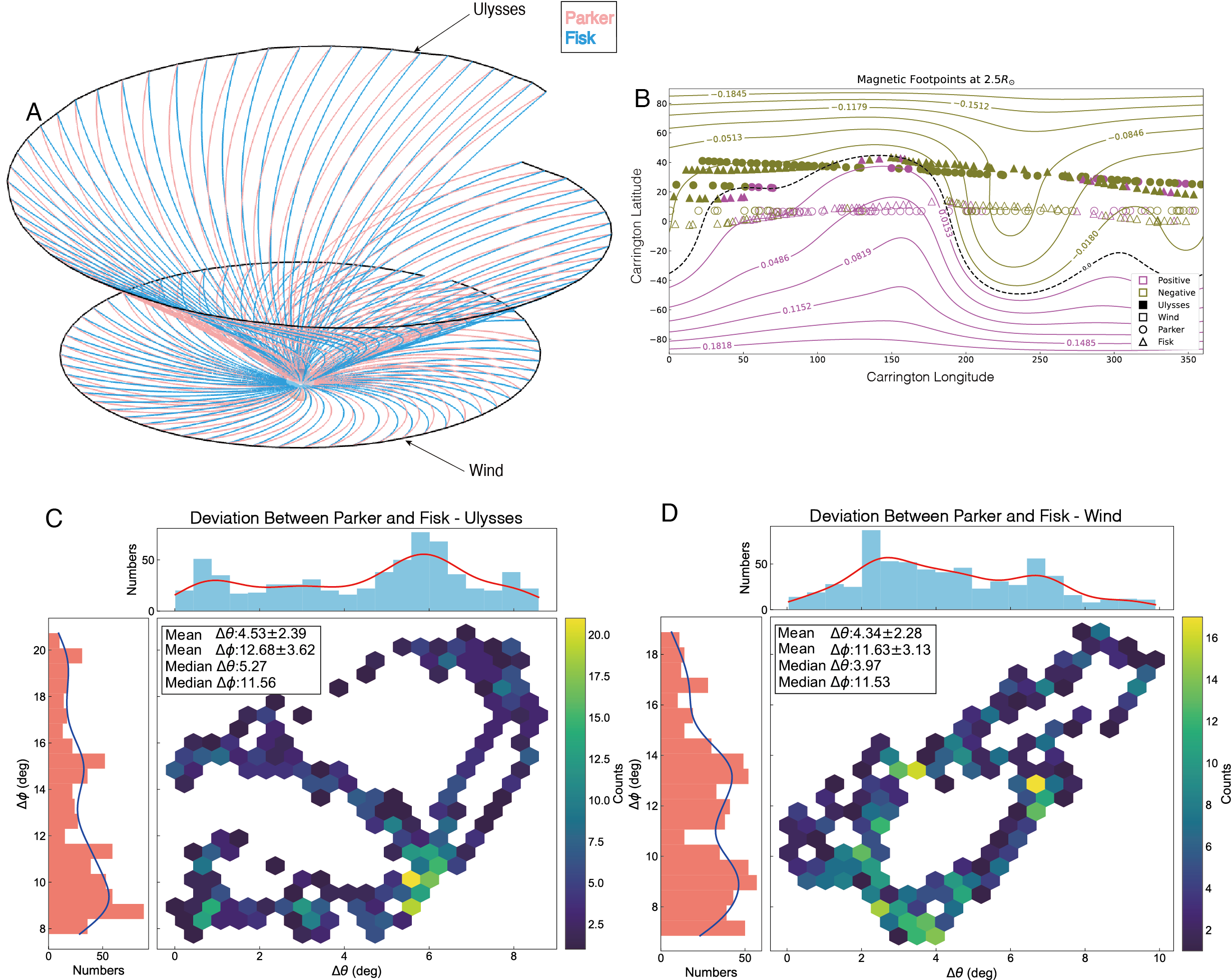}
  \caption{Backmapping of the solar wind measured by the Ulysses and Wind spacecraft to the source surface of the PFSS model ($2.5 R_{\odot}$) based on the Parker and Fisk HMF model.
  Panel A: examples of Parker (red) and Fisk (blue) magnetic field lines traced from identical measurement points to $r=2.5 R_{\odot}$.
  Panel B: magnetic footpoints at $2.5R_{\odot}$ traced from measurements by Ulysses (solid markers) and Wind (hollow markers) using the Parker (circles) and Fisk (triangles) models. Color coding indicates polarity: olive for negative, magenta for positive, and black dotted lines for neutral line.
  Panel C: latitudinal ($\Delta\theta$) and longitudinal ($\Delta\phi$) deviations between Parker and Fisk magnetic footpoints at $r=2.5 R_{\odot}$, traced from Ulysses measurements. The color encodes the count. The blue and red lines overlaid on the histograms represent the smoothed probability density distributions of the deviation values, derived from Kernel Density Estimation (KDE).
  Panel D: the same as Panel C but for Wind measurements.}
  \label{fig:2}
\end{figure*}
Figure \ref{fig:2} presents the backmapping of the solar wind measured by the Ulysses and Wind spacecraft to the source surface of the PFSS model, based on the Parker and Fisk HMF models. Figure \ref{fig:2}(A) shows examples of Parker (red) and Fisk (blue) magnetic field lines traced from identical measurement points to $r=2.5 R_{\odot}$. Although both the Parker and Fisk fields form spirals around the Sun, the Fisk field diverges a little from the Parker field.

In Figure \ref{fig:2}(B), we exhibit magnetic footpoints at $2.5R_{\odot}$ traced from measurements by Ulysses (solid markers) and Wind (hollow markers) using the Parker (circles) and Fisk (triangles) models. Polarity is color-coded: olive for negative, magenta for positive, with the neutral line indicated by a black dotted line.
Figure \ref{fig:2}(B) reveals that the Ulysses measurements map back to regions of negative polarity. In contrast, the Wind measurements are divided by the large-scale heliospheric current sheet (HCS), placing one part in positive polarity regions and the other in negative polarity regions. Furthermore, magnetic field footpoints at $2.5 R_{\odot}$ connected to Ulysses and Wind by the Parker fields exhibit slight deviations from those mapped using the Fisk fields.

We quantify the deviations of magnetic field footpoints at $2.5 R_{\odot}$ between the Parker and Fisk fields in Figure \ref{fig:2}(C) and (D), which show that for Ulysses (mid-latitude) and Wind (ecliptic plane) measurements, the latitudinal differences between magnetic footpoint positions obtained by Parker and Fisk fields range from $\Delta \theta = 0^\circ$ to $\Delta \theta = \sim 10 ^\circ$. The mean values are $4.53^\circ\pm 2.39^\circ$ (Ulysses) and $4.34^\circ\pm 2.28^\circ$ (Wind), with median values being $5.27^\circ$ and $3.97^\circ$, respectively.
The longitudinal differences exhibit a broader spread from $\Delta \phi = 0^\circ$ to $\Delta \phi = \sim 20 ^\circ$, with averages of $12.68^\circ\pm 3.62^\circ$ (Ulysses) and $11.63^\circ\pm 3.13^\circ$ (Wind), and median values of $11.56^\circ$ (Ulysses) and $11.53^\circ$ (Wind).

\begin{figure*}[htbp]
  \centering
  \includegraphics[width=1\textwidth]{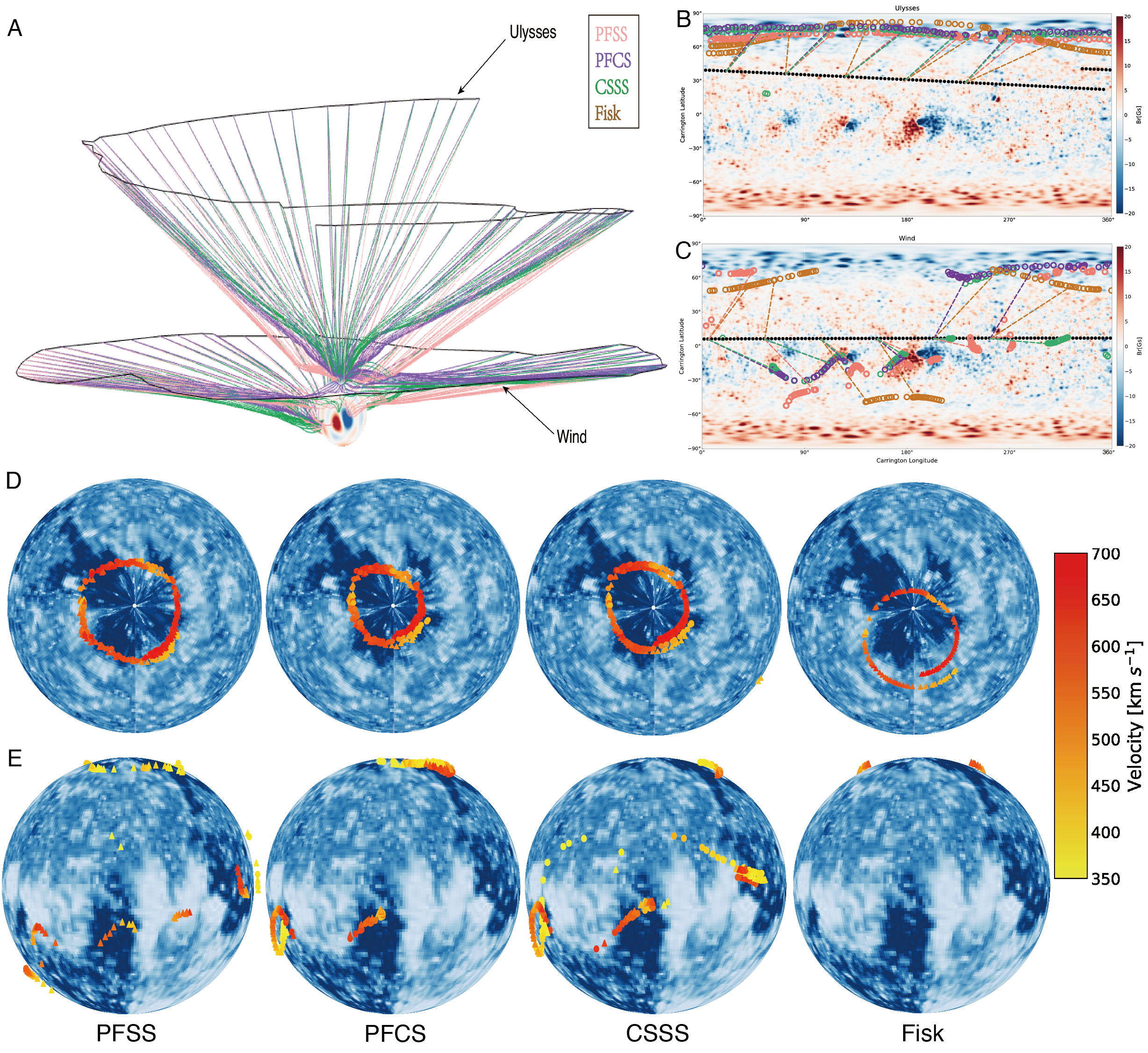}
  \caption{Backmapping of the solar wind measured by the Ulysses and Wind spacecraft from the source surface ($2.5 R_{\odot}$) to the photosphere based on the Fisk HMF model.
  Panel A: representative magnetic field lines from the PFSS (pink), PFCS (darkviolet), and CSSS (mediumseagreen) CMF models.
  Panels B and C: corresponding photospheric footpoints traced from the Ulysses (B) and Wind (C) in situ observations, overlaid on a synoptic magnetogram, respectively.
  Panel D: A view of the Sun's north pole created by reprojecting the STEREO-A/SECCHI EUVI 195 $\mathring{\mathrm{A}}$ synoptic map onto a spherical coordinate system, with backmapped solar wind source regions color-coded by speed during Ulysses' mid-latitude scan.
  Panel E: the same as Panel D but for Wind measurements.}
  \label{fig:3}
\end{figure*}

Following a comparative analysis of the Parker and Fisk HMF models, we utilize the boundary footpoints derived from the Fisk HMF model within the CMF model for subsequent backmapping to the solar surface. The results are presented in Figure \ref{fig:3}.
Figure \ref{fig:3}(A) illustrates representative magnetic field lines from the PFSS (pink), PFCS (darkviolet), and CSSS (mediumseagreen) models.
Figure \ref{fig:3}(B) and (C) show the corresponding photospheric footpoints traced from the Ulysses and Wind in situ observations, overlaid on a synoptic magnetogram, respectively.

The magnetic field line connectivity predicted by the CMF models exhibits significant variation. For Ulysses mid-latitude measurements, most solar wind maps to high latitudes. While the footpoints obtained using the PFSS, PFCS, and CSSS models converge spatially, those predicted by the Fisk model deviate evidently from the PFSS, PFCS, and CSSS models.

For Wind ecliptic measurements, the mapping results are more complex, with all four models predicting markedly different connectivity. In some cases, the PFSS and Fisk models map the solar wind to sources in the northern hemisphere, whereas the PFCS and CSSS models map the same wind to the southern hemisphere. Furthermore, instances occur where the PFSS, PFCS, and CSSS models map to the same active region, yet connect to footpoints of opposing magnetic polarities.

Figure \ref{fig:3}(D) displays a view of the Sun's north pole created by reprojecting the STEREO-A/SECCHI EUVI 195 $\mathring{\mathrm{A}}$ synoptic map onto a spherical coordinate system, alongside backmapped solar wind source regions color-coded by speed during Ulysses' mid-latitude scan.
Coronal holes, which are characterized by relatively low EUV emissions due to reduced plasma density, constitute the primary source of the Sun's open magnetic field \citep{1977Zirker}. All four CMF models predict that the fast solar wind measured by Ulysses originates from polar coronal holes. This aligns with established research indicating that the fast solar wind ($>$ 700 km s$^{-1}$) originates from polar coronal holes \citep{1995PHILLIPS,1998Neugebauer,2005Lionello,2021Reiss}.

\begin{figure*}[htbp]
  \centering
  \includegraphics[width=1\textwidth]{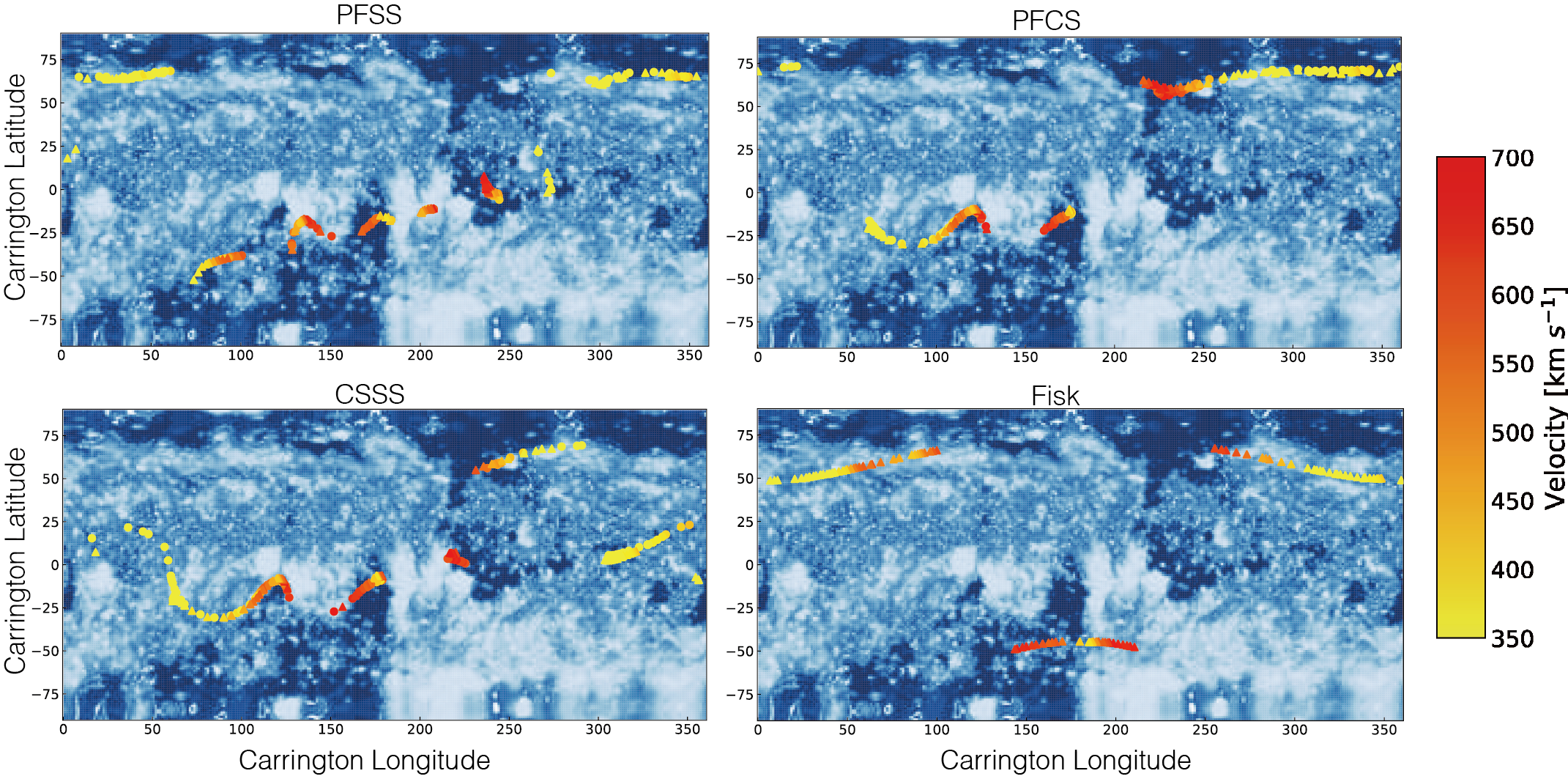}
  \caption{Backmapped source regions of the Wind-measured solar wind superimposed on an  EUVI synoptic image.}
  \label{fig:4}
\end{figure*}

Figure \ref{fig:3}(E) presents backmapped source regions of the Wind-measured solar wind superimposed on EUV full-disk images. To further contextualize these source regions, Figure \ref{fig:4} shows corresponding footpoints overlaid on an EUVI synoptic map.
The PFSS, PFCS, and CSSS models map portions of the solar wind to both mid/low-latitude coronal holes and the periphery of the north polar coronal hole. Notably, the PFSS model associates fast solar wind ($>$ 500 km s$^{-1}$) with low-latitude coronal holes or active regions, whereas PFCS and CSSS models connect some fast wind streams to the polar coronal hole boundary. In contrast, the Fisk model uniquely predicts some of low-latitude solar wind sources within quiet-Sun regions exhibiting high EUV emissions.

\begin{figure*}[htbp]
  \centering
  \includegraphics[width=1\textwidth]{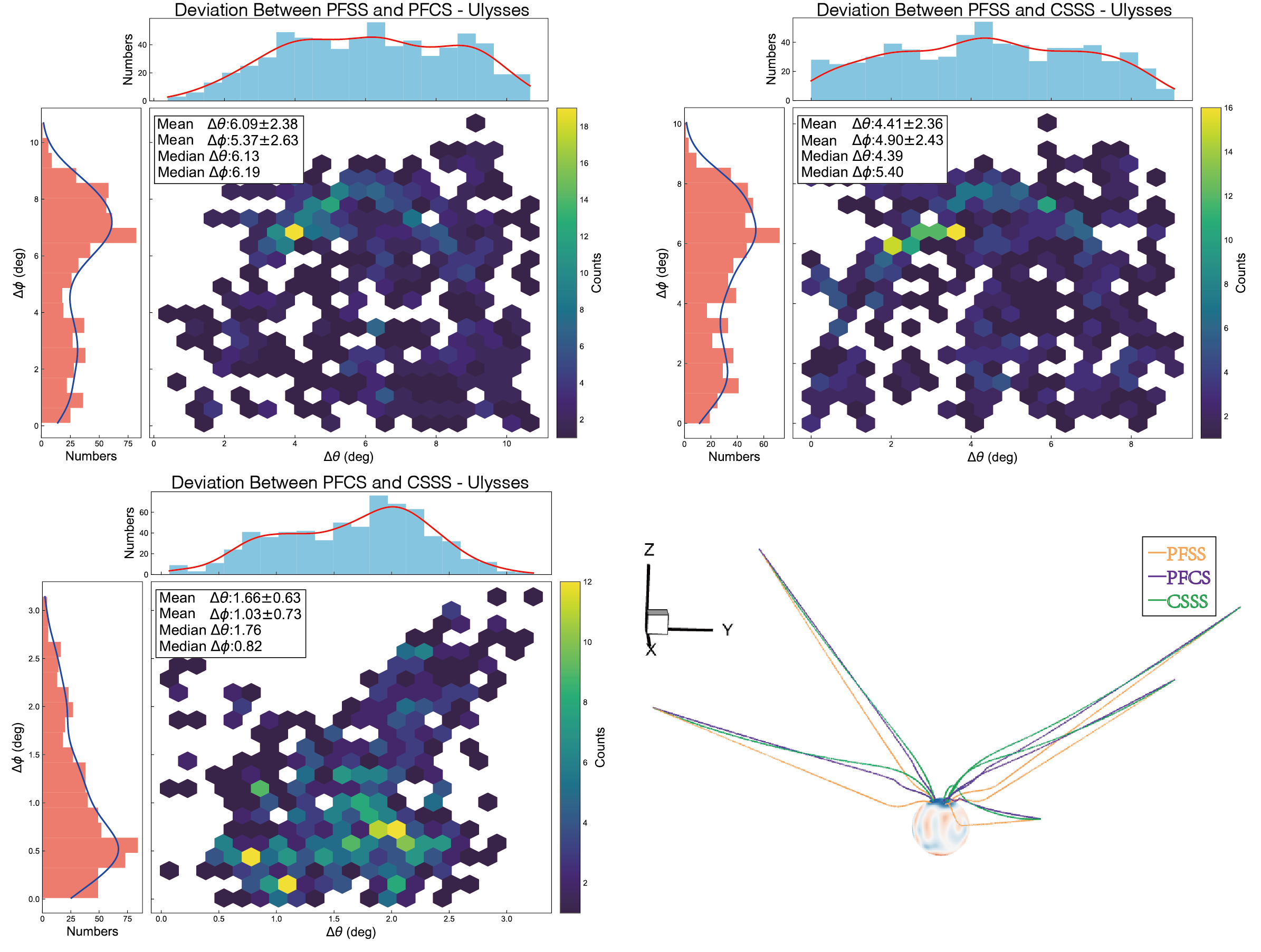}
  \caption{Latitudinal ($\Delta\theta$) and longitudinal ($\Delta\phi$) deviations of photospheric footpoints among the PFSS, PFCS, and CSSS models for Ulysses measurements. Representative magnetic field lines traced from identical points at the model outer boundaries to the photosphere are included for contextual comparison.}
  \label{fig:5}
\end{figure*}

\begin{figure*}[htbp]
  \centering
  \includegraphics[width=1\textwidth]{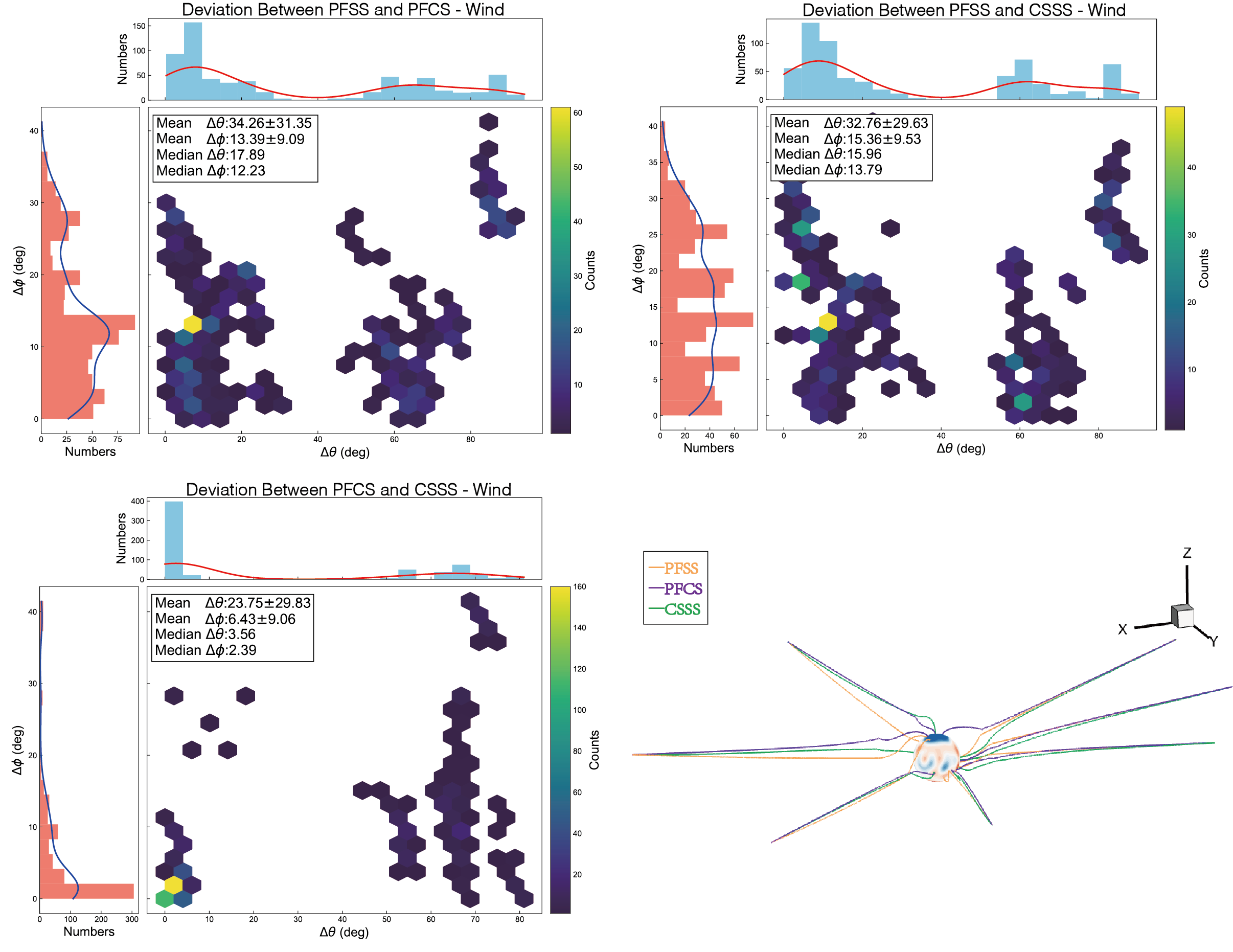}
  \caption{The same as Figure \ref{fig:5} but for Wind measurements.}
  \label{fig:6}
\end{figure*}

Figures \ref{fig:5} and \ref{fig:6} present latitudinal ($\Delta\theta$) and longitudinal ($\Delta\phi$) deviations of photospheric footpoints among the PFSS, PFCS, and CSSS models for Ulysses and Wind measurements, respectively. Representative magnetic field lines traced from identical points at the model outer boundaries to the photosphere are included for contextual comparison.

For Ulysses mid-latitude measurements, angular deviations between models are minimal. The closest agreement occurs between the PFCS and CSSS models, with mean deviations of $1.66^\circ \pm 0.63^\circ$ in latitude and $1.03^\circ \pm 0.73^\circ$ in longitude, and median deviations of $1.76^\circ$ and $0.82^\circ$, respectively. Larger but moderate differences are observed between other pairs: mean $\Delta\theta = 6.09^\circ\pm 2.38^\circ$ (median: $6.13^\circ$), mean $\Delta\phi = 5.37^\circ\pm 2.63^\circ$ (median: $6.19^\circ$) between the PFSS and PFCS models, and mean $\Delta\theta = 4.41^\circ\pm 2.36^\circ$ (median: $4.39^\circ$), mean $\Delta\phi = 4.90^\circ\pm 2.43^\circ$ (median: $5.40^\circ$) between the PFSS and CSSS models.

In contrast, Wind ecliptic measurements exhibit substantial model divergence, with maximum deviations reaching up to $40^\circ$ ($\Delta\phi$) and $80^\circ$ ($\Delta\theta$). Latitudinal differences systematically exceed longitudinal variations across all model comparisons. Notably, PFCS-CSSS pairs maintain minimal separation, consistent with mid-latitude backmapping trends.

\subsection{High- and Low-latitude Backmapping}
We extend our validation to CR 2060 (low-latitude focused, 14 August - 10 September 2007) and CR 2064 (high-latitude focused, 1-29 December 2007), maintaining identical methodology as previously defined in the ecliptic and middle latitudes.

\begin{figure*}[htbp]
  \centering
  \includegraphics[width=1\textwidth]{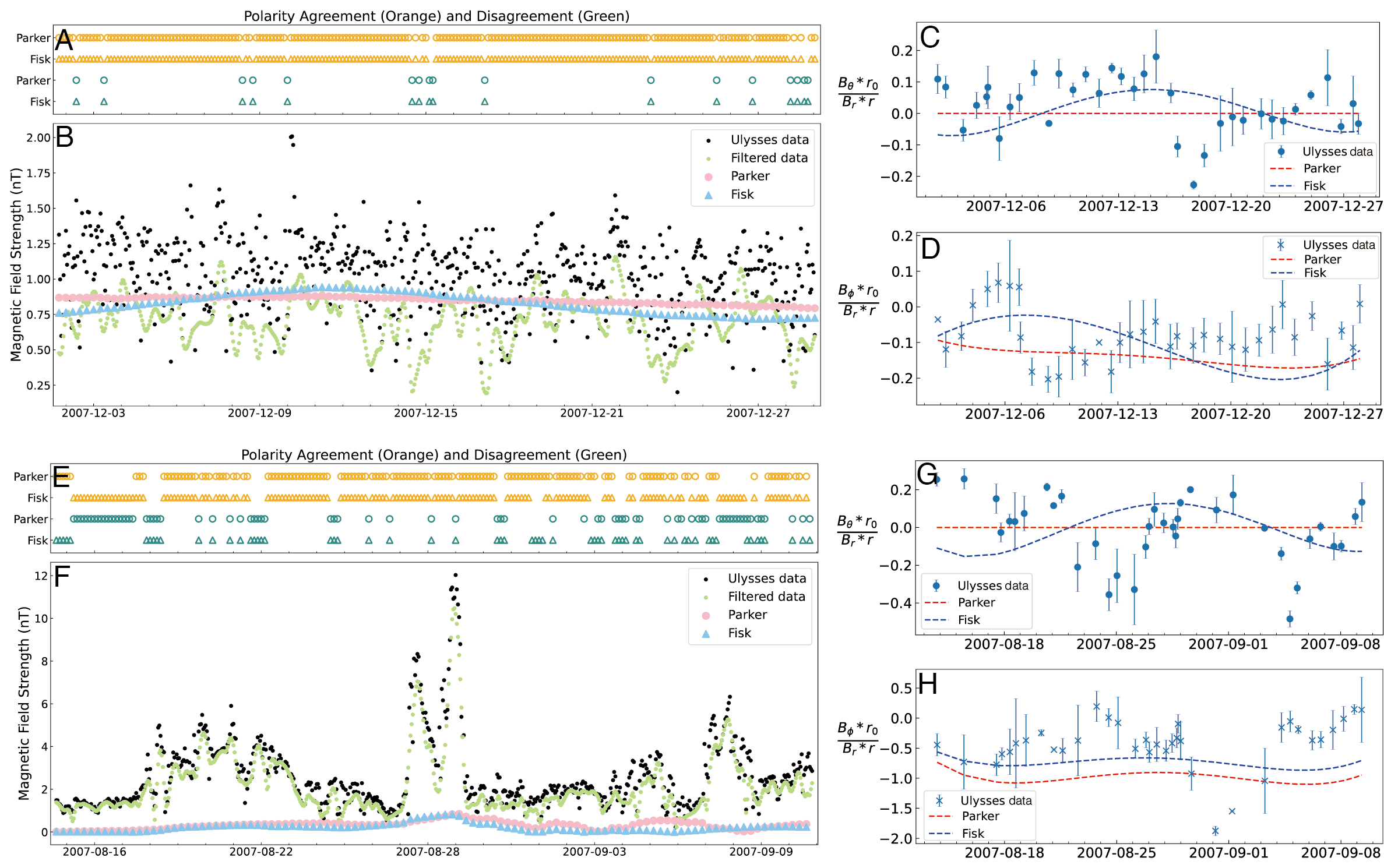}
  \caption{The same as Figure \ref{fig:1} but for high- (A, B, C, D) and low-latitude (E, F, G, H) measurements by Ulysses.}
  \label{fig:7}
\end{figure*}

Figure \ref{fig:7} evaluates the Parker and Fisk HMF model predictions of HMFs' polarity, strength, and configuration against Ulysses measurements at both high and low latitudes. Figure \ref{fig:7}(A) and (B) demonstrate strong agreement at high latitudes, with both models achieving $92.52\%$ polarity match rates with observations. Residual field strengths remain minimal (Parker: 0.18 nT; Fisk: 0.19 nT), indicating excellent consistency between predicted and observed magnitudes. As quantified by the configuration metrics $B_\theta r_0 / (B_r r)$ and $B_\phi r_0 / (B_r r)$ in Figure \ref{fig:7}(C) and (D), the Fisk model successfully captures the global trend of high-latitude measured magnetic topology despite localized deviations.

Figure \ref{fig:7} panels (E-H) evaluate Parker and Fisk HMF model performance against Ulysses low-latitude measurements, with comparative metrics benchmarked to Wind ecliptic measurements (corresponding to Figure \ref{fig:1}(E-H)). The Parker model achieves $65.60\%$ polarity agreement with Ulysses observations, while the Fisk model shows a significant improvement, reaching $72.48\%$. The residual field strengths is 1.84 nT for the Parker fields and 1.94 nT for the Fisk fields. While neither model fully captures the complexity of low-latitude field configurations, the Fisk model demonstrates comparatively better alignment with the Ulysses low-latitude measurements.

\begin{figure*}[htbp]
  \centering
  \includegraphics[width=1\textwidth]{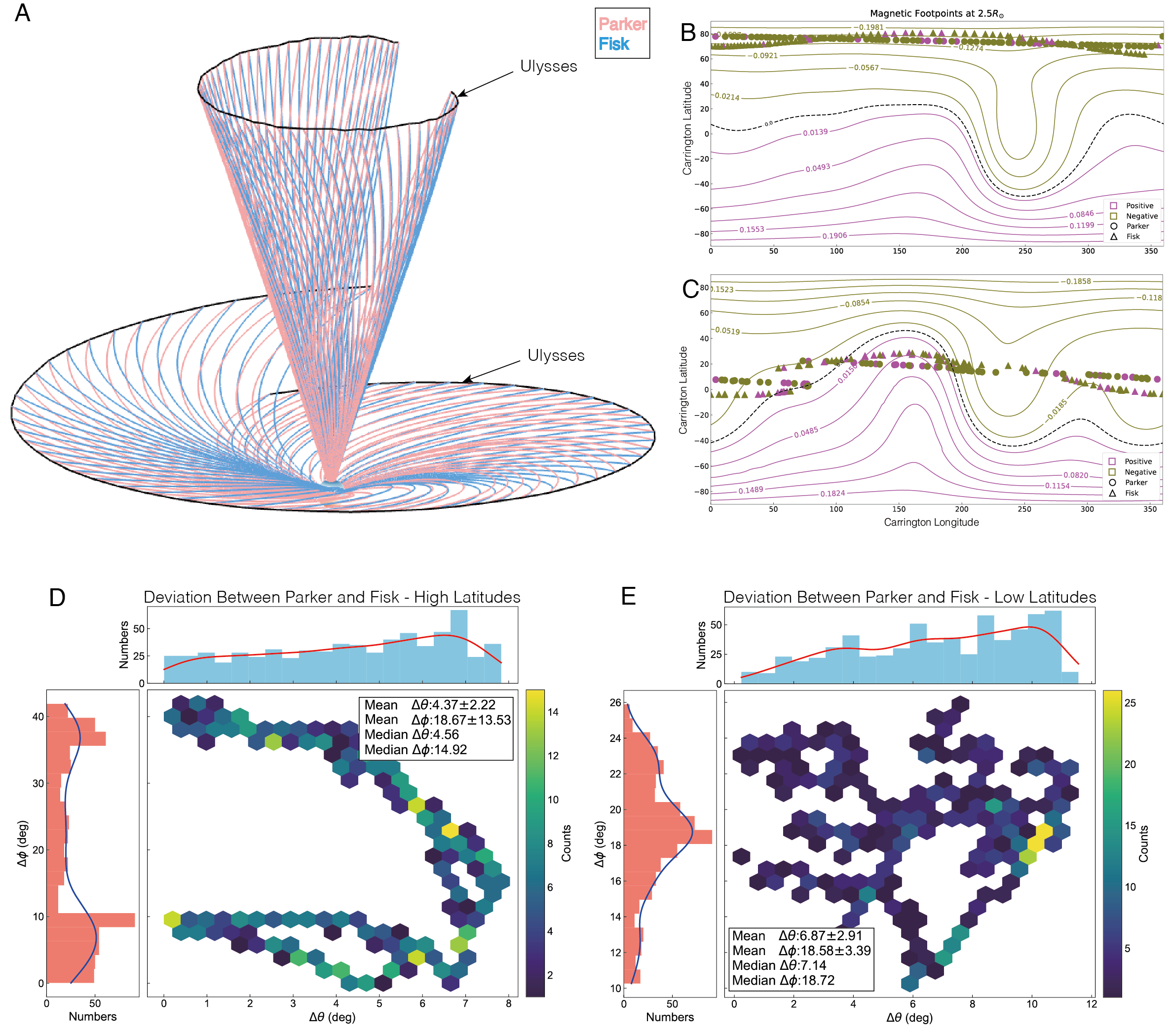}
  \caption{The same as Figure \ref{fig:2} but for high- and low-latitude measurements by Ulysses.}
  \label{fig:8}
\end{figure*}

Figure \ref{fig:8} compares backmapped footpoint locations at the PFSS source surface ($2.5 R_{\odot}$) for Ulysses high- and low-latitude measurements using the Parker versus Fisk HMF models. Although both models predict similar large-scale magnetic topologies, their derived footpoints exhibit significant spatial separations across for the Ulysses high- and low-latitude measurements. These separations show isotropic scattering across the $\Delta\theta$-$\Delta\phi$ plane. The mean longitudinal deviation between the Parker and Fisk HMF models is as high as $18.67^\circ \pm 13.53^\circ$ for the high-latitude measurements, and $18.58^\circ \pm 3.39^\circ$ for the low-latitude measurements, with median longitudinal deviation being $14.92^\circ$ and $18.72^\circ$, respectively. While latitudinal offsets remain consistently smaller than longitudinal deviations for both regimes, they still reach mean (median) values of several degrees.

\begin{figure*}[htbp]
  \centering
  \includegraphics[width=1\textwidth]{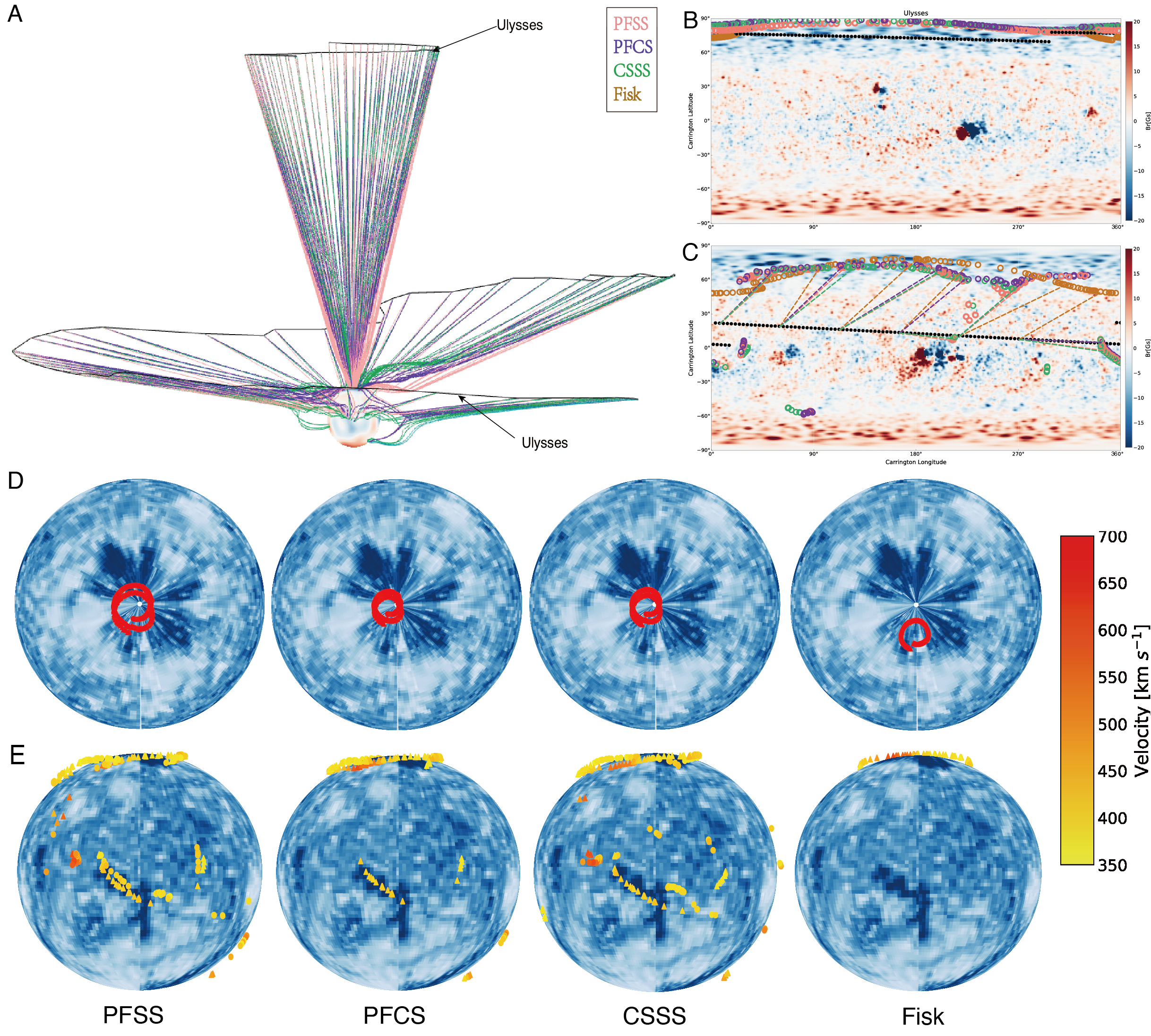}
  \caption{The same as Figure \ref{fig:3} but for high- and low-latitude measurements by Ulysses.}
  \label{fig:9}
\end{figure*}

Figure \ref{fig:9} presents backmapped source regions of Ulysses high- and low-latitude solar wind using the PFSS, PFCS, CSSS, and Fisk CMF models. For high-latitude measurements, the PFSS, PFCS, and CSSS models show minimal field line divergence, with most photospheric footpoints clustered near the latitude of $85^{\circ}$. EUVI imagery in Figure \ref{fig:9} confirms these models map the fast solar wind($>$ 750 km s$^{-1}$) back to the central region of the northern polar coronal hole. The Fisk model places footpoints farther from the pole (consistent with its underlying assumptions), with some even extending to the boundary of the coronal hole.

For Ulysses low-latitude measurements, magnetic connectivity exhibits significant divergence among PFSS, PFCS, and CSSS models, resulting in substantial spatial separation of photospheric footpoints across all three solutions as shown in Figure \ref{fig:9}(C). These conventional models map most low-speed solar wind ($<$ 500 km s$^{-1}$) to the periphery of the northern polar coronal hole as shown in Figure \ref{fig:9}(E). Notably, the PFSS, PFCS, and CSSS models associate some low-speed wind with isolated low-latitude coronal holes, while the Fisk model exclusively links all such wind to polar coronal holes.

\begin{figure*}[htbp]
  \centering
  \includegraphics[width=1\textwidth]{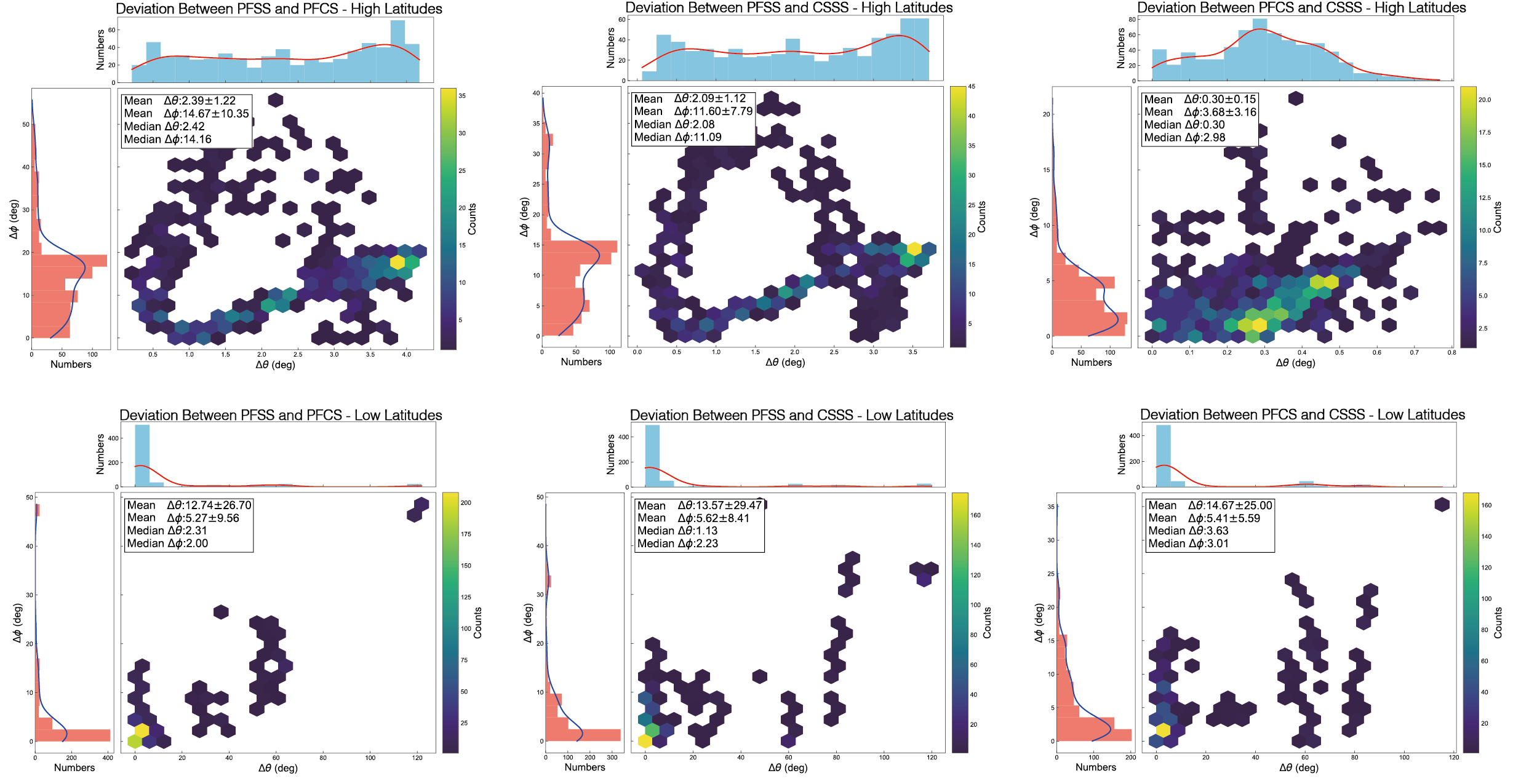}
  \caption{Latitudinal ($\Delta\theta$) and longitudinal ($\Delta\phi$) deviations of photospheric footpoints among the PFSS, PFCS, and CSSS models for Ulysses high-(up panels) and low-latitude (down panels) measurements.}
  \label{fig:10}
\end{figure*}

Figure \ref{fig:10} presents the latitudinal ($\Delta\theta$) and longitudinal ($\Delta\phi$) deviations of photospheric footpoints among the PFSS, PFCS, and CSSS models for Ulysses high- and low-latitude measurements.
For Ulysses high-latitude measurements, the angular separations between the models are widely scattered across the $\Delta\theta$-$\Delta\phi$ plane. The mean and median $\Delta\theta$ values are significantly smaller than those of $\Delta\phi$. Similarly, the PFCS and CSSS models exhibit the closest agreement. For low-latitude measurements, the angular separations predominantly cluster within a narrow range of small $\Delta\theta$ and $\Delta\phi$ values. However, some cases exist where deviations between the models reach $50^\circ$ in longitude ($\Delta\phi$) and $120^\circ$ in latitude ($\Delta\theta$).

\subsection{Backmapping with Different Source Surface Heights}

In the PFSS CMF model, the source surface height $R_\mathrm{{SS}}$ represents a critical yet empirically constrained parameter that can significantly influences reconstructed coronal magnetic field topologies and subsequent solar wind predictions \citep{2025Koukras}. Adjusting $R_\mathrm{{SS}}$ can directly modulates the extent of closed fields and the distribution of open fields. This sensitivity necessitates careful optimization of $R_\mathrm{{SS}}$ to reconcile model outputs with in situ solar wind measurements and remote-sensing observations (e.g., white-light coronagraphs or EUV coronal hole boundaries). Here, we investigate how variations in $R_\mathrm{{SS}}$ affect the identification of solar wind source regions.

\begin{figure*}[htbp]
  \centering
  \includegraphics[width=1\textwidth]{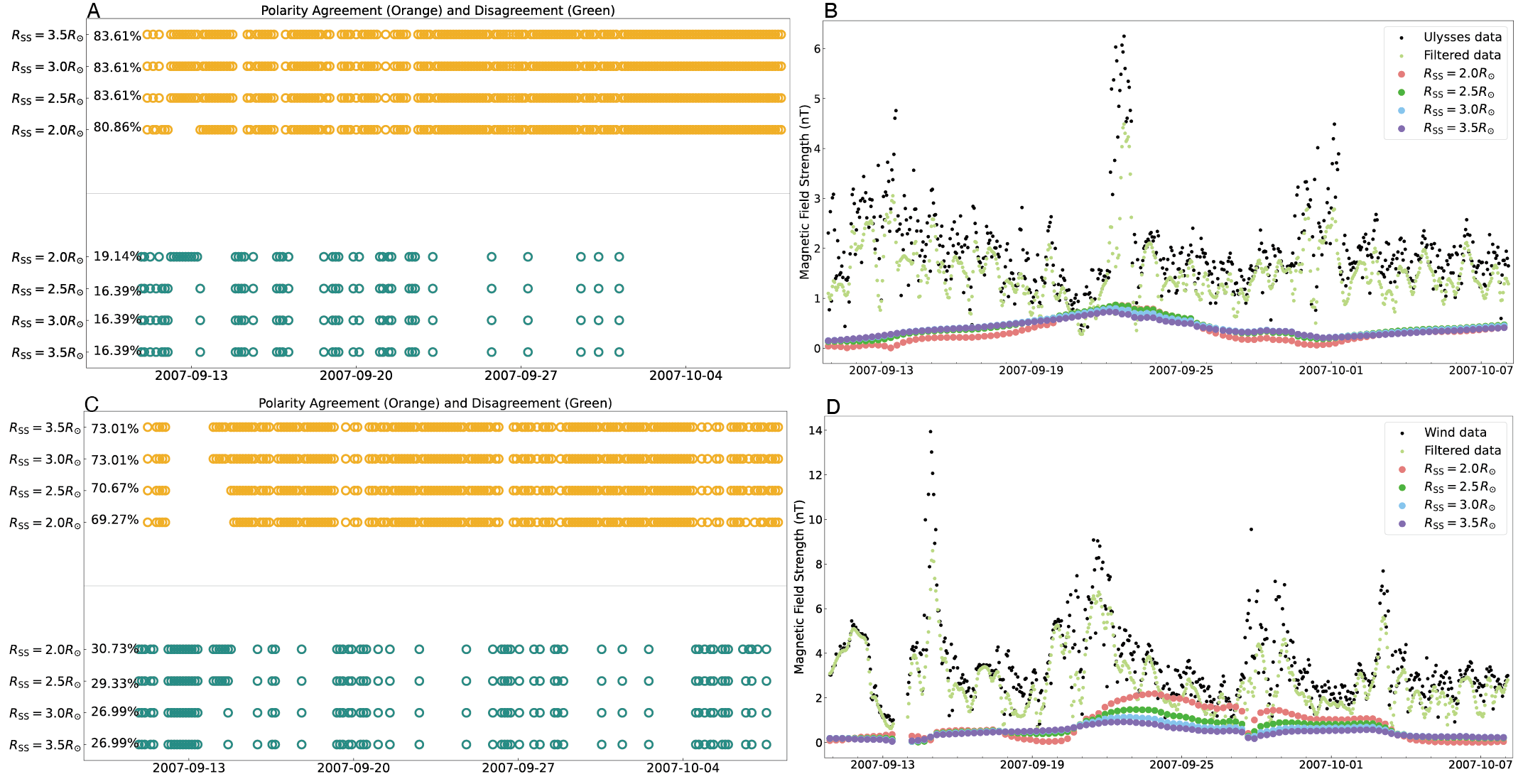}
  \caption{Observed and predicted HMF polarity and strength with in situ measurements obtained from Ulysses and Wind.}
  \label{fig:11}
\end{figure*}

Figure \ref{fig:11} compares observed and predicted HMF polarity and strength using mid-latitude and ecliptic measurements from Ulysses and Wind. The models employed are the PFSS CMF model coupled with Parker's HMF model, tested at source surface heights of $R_\mathrm{SS}=2.0R_{\odot}$, $R_\mathrm{SS}=2.5R_{\odot}$, $R_\mathrm{SS}=3.0R_{\odot}$ and $R_\mathrm{SS}=3.5R_{\odot}$.
Variations in $R_\mathrm{SS}$ have limited impact on mid-latitude HMF predictions: all four cases achieve similar polarity match rates and residual field strengths relative to observations.
For ecliptic HMFs, polarity match rates remain comparable across $R_\mathrm{SS}$ values, while residual field magnitudes increase from 1.97 nT for $R_\mathrm{SS}=2.0R_{\odot}$ to 2.26 nT for $R_\mathrm{SS}=3.5R_{\odot}$.
Consistent with this trend, lower $R_\mathrm{SS}$ values yield stronger ecliptic HMFs.

\begin{figure*}[htbp]
  \centering
  \includegraphics[width=1\textwidth]{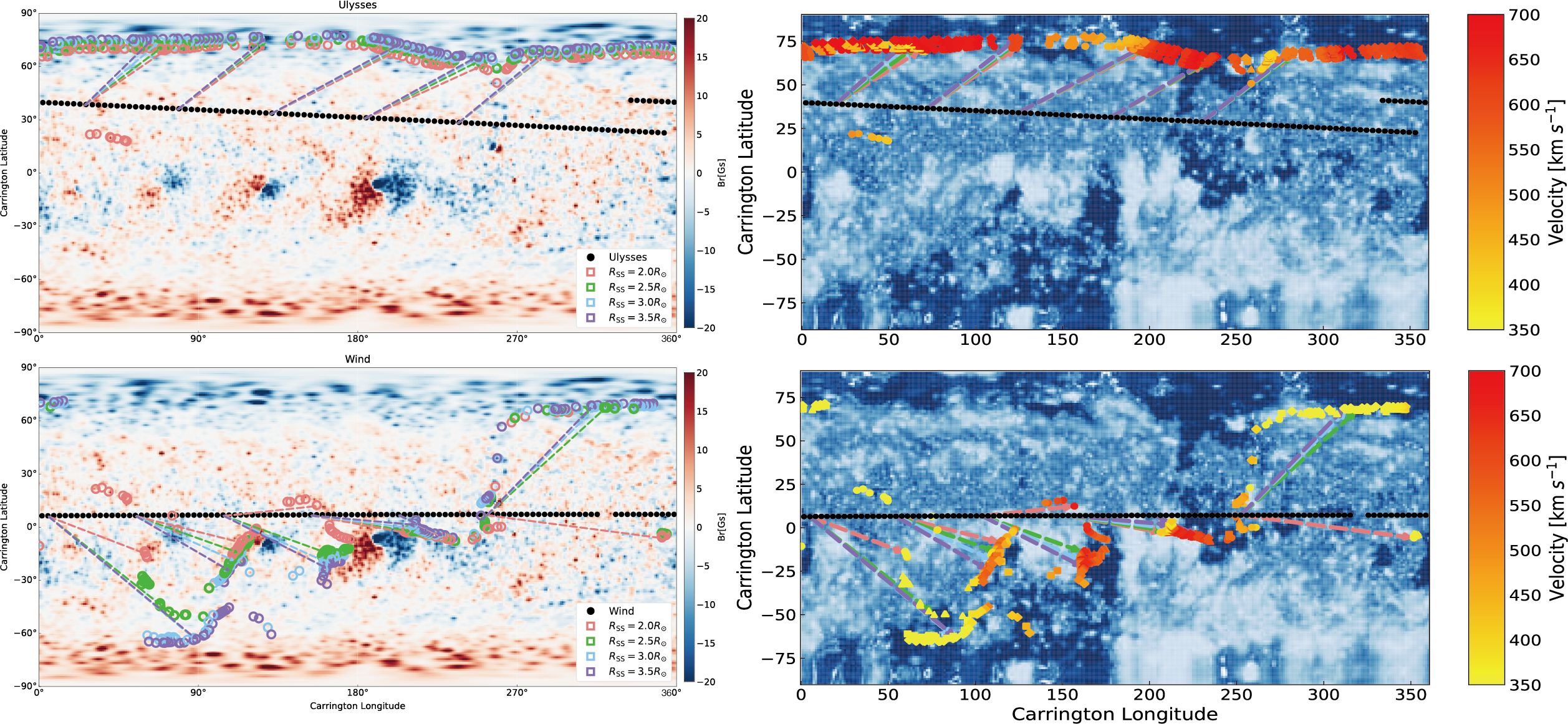}
  \caption{Photospheric footpoints traced from the different source surface of the PFSS model on the GONG synoptic magnetogram and EUVI synoptic map.}
  \label{fig:12}
\end{figure*}

Figure \ref{fig:12} shows photospheric footpoints locations derived from Ulysses mid-latitude and Wind ecliptic measurements according to the different heights of the source surface. The left column overlays the footpoints on GONG synoptic magnetograms, while the right column uses EUVI synoptic maps. 
Variations in the source surface height induce more pronounced changes in footpoint locations for ecliptic measurements than for mid-latitude observations. For the ecliptic solar wind, the PFSS model with $R_\mathrm{SS}=2.0R_{\odot}$ maps some sources to mid-latitudes, while these same sources shift to high latitudes by the other three cases. The identical ecliptic solar wind is mapped back to the quiet-Sun region with $R_\mathrm{SS}=2.0R_{\odot}$, and back to the active region with $R_\mathrm{SS}=2.5R_{\odot}$, and to a mid-latitude coronal hole with $R_\mathrm{SS}=3.0R_{\odot}$ and $R_\mathrm{SS}=3.5R_{\odot}$. Mid-latitude solar wind sources consistently map near polar coronal hole boundaries across all $R_\mathrm{SS}$ values. However, at $R_\mathrm{SS}=2.0R_{\odot}$, a subset shifts to low-latitude quiet-Sun regions.

\begin{figure*}[htbp]
  \centering
  \includegraphics[width=1\textwidth]{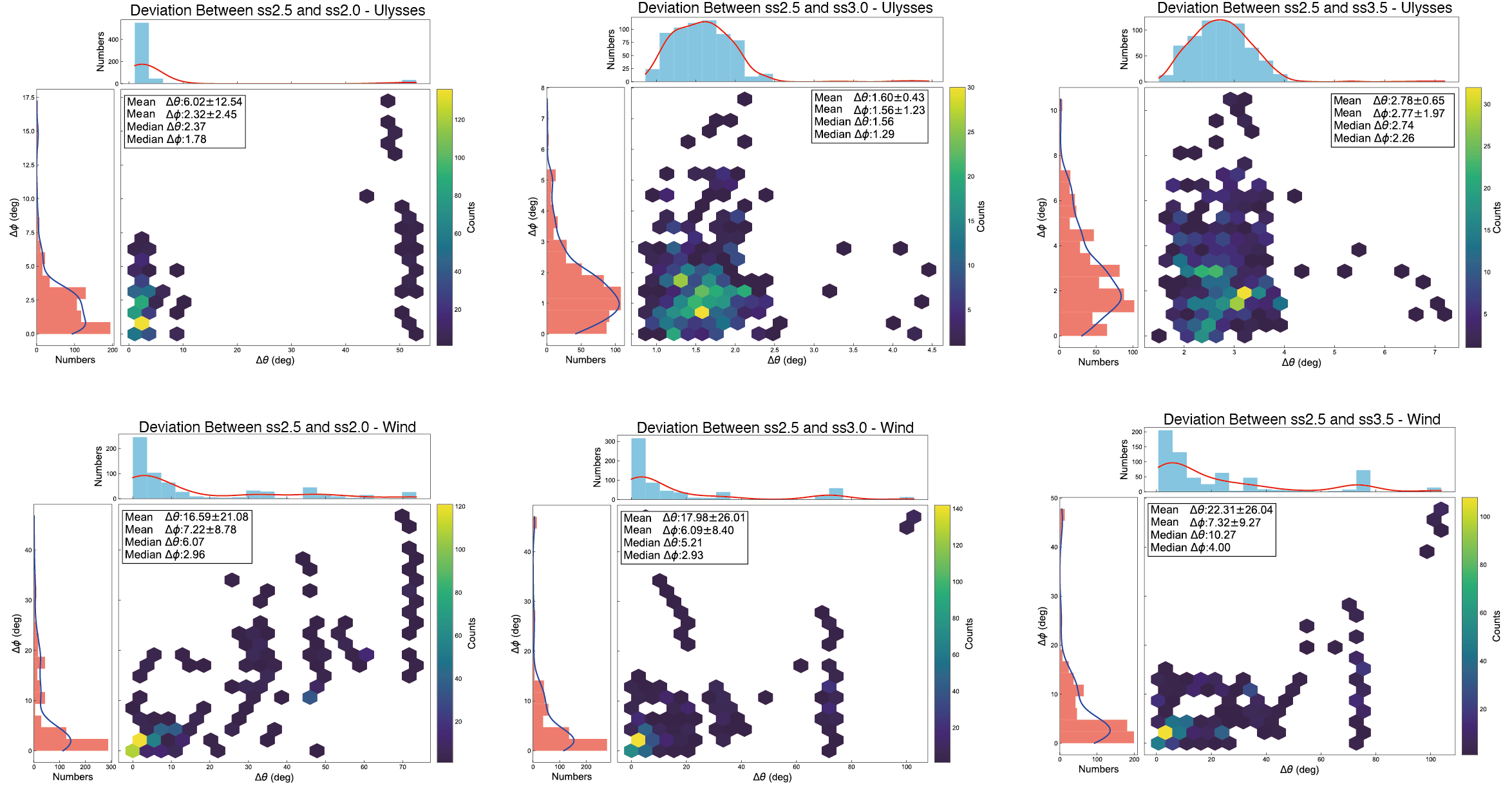}
  \caption{Latitudinal ($\Delta\theta$) and longitudinal ($\Delta\phi$) deviations of photospheric footpoints among the different source surface of the PFSS model for Ulysses mid-latitude (up panels) and Wind ecliptic (down panels) measurements.}
  \label{fig:13}
\end{figure*}

Figure \ref{fig:13} presents latitudinal ($\Delta\theta$) and longitudinal ($\Delta\phi$) deviations of photospheric footpoints among the different source surface heights for Ulysses mid-latitude (up panels) and Wind ecliptic (down panels) measurements.
For Ulysses mid-latitude measurements, footpoints converge toward consistent locations at $R_\mathrm{SS}\geq 2.5R_{\odot}$.
The mean (median) deviation between successive heights is $1.60^\circ \pm 0.43^\circ$ ($1.56^\circ$) in latitude and $1.56^\circ \pm 1.23^\circ$ ($1.29^\circ$) in longitude between $R_\mathrm{SS}=2.5R_{\odot}$ and $R_\mathrm{SS}=3.0R_{\odot}$, and is $2.78^\circ \pm 0.65^\circ$ ($2.74^\circ$) in latitude and $2.77^\circ \pm 1.97^\circ$ ($2.26^\circ$) in longitude between $R_\mathrm{SS}=2.5R_{\odot}$ and $R_\mathrm{SS}=3.5R_{\odot}$. Although large differences exist between $R_\mathrm{SS}=2.0R_{\odot}$ and $R_\mathrm{SS}=2.5R_{\odot}$, angular separations remain confined within narrow $\Delta\theta$-$\Delta\phi$ ranges.
For Wind ecliptic measurements, deviations between $R_{\mathrm{SS}}$ pairs (e.g., $2.0R_{\odot}$ vs. $2.5R_{\odot}$) exhibit bimodal distribution: predominantly small angular differences with occasional large separations ($>50^\circ$).

\section{Conclusions and Discussion} \label{sec:conclu}

In this study, we combined the Parker and Fisk HMF models with the PFSS, PFCS, and CSSS CMF models to trace the solar wind back to the solar disk. Our analysis, leveraging in situ Ulysses measurements across latitudes, GONG ADAPT synoptic magnetograms, STEREO-A EUVI observations, demonstrates that the specific configurations of both the CMF and HMF models significantly influence the identification of solar wind source regions. Crucially, we quantified angular deviations in the backmapping results arising from different CMF/HMF combinations, with particular implications for accurately pinpointing the origins of fast solar wind emanating from the Sun's polar regions.

\begin{table}[htbp]
  \centering
  \caption{Difference Between the Parker and Fisk HMF Models in Prediction and Footpoint Positions.}
  \includegraphics[width=1\textwidth]{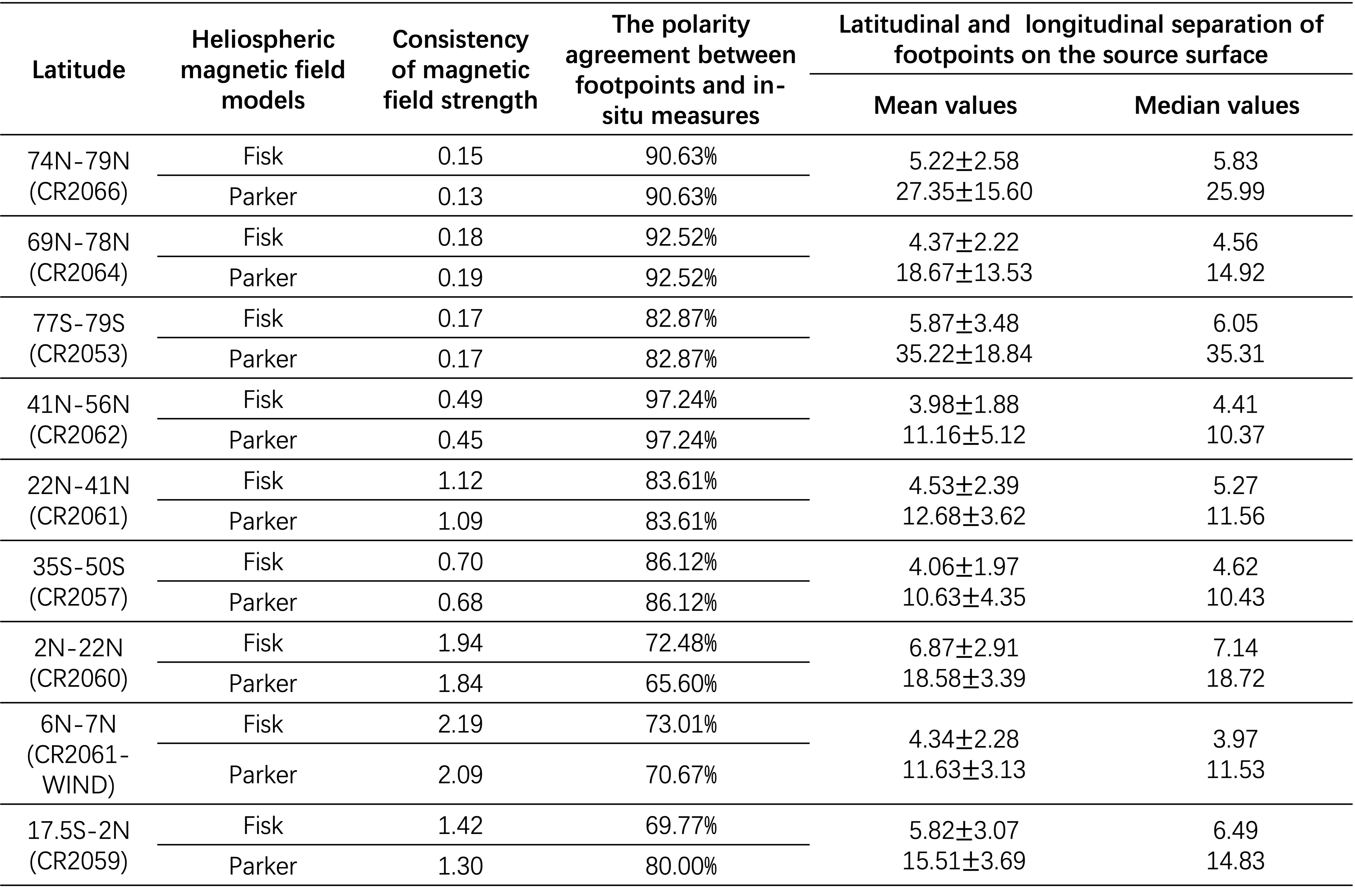}
  \label{tab:1}
\end{table}

Our analysis reveals performance differences between the Parker and Fisk HMF models across latitudes (see Table \ref{tab:1}). Both models achieve similar polarity match rates and residual field strengths relative to Ulysses measurements, with the higher consistency between predicted and measured HMFs at high latitudes than at low latitudes. Although the Fisk HMF model improves upon the Parker model in representing the meridional field component, it is unable to fully capture the complex structure of the HMF across the full latitudinal range. Critically, when tracing solar wind measurements back to the coronal source surface ($2.5 R_{\odot}$), the Fisk HMF model produces significantly displaced magnetic footpoints compared to the Parker model, despite similar overall HMF configurations. Latitudinal offsets between the two models' footpoints average several degrees, with a comparable median value, confirming that these offsets are substantially smaller than the longitudinal deviations. Furthermore, the longitudinal deviation between the Parker and Fisk models exhibits a clear positive trend with increasing latitude, a feature that is robustly captured by both the rising mean values and the closely aligned median values.

\begin{table}[htbp]
  \centering
  \caption{Separation and Position of Photospheric Footpoints among the PFSS, PFCS, and CSSS CMF Models under the Fisk HMF Model.}
  \includegraphics[width=1\textwidth]{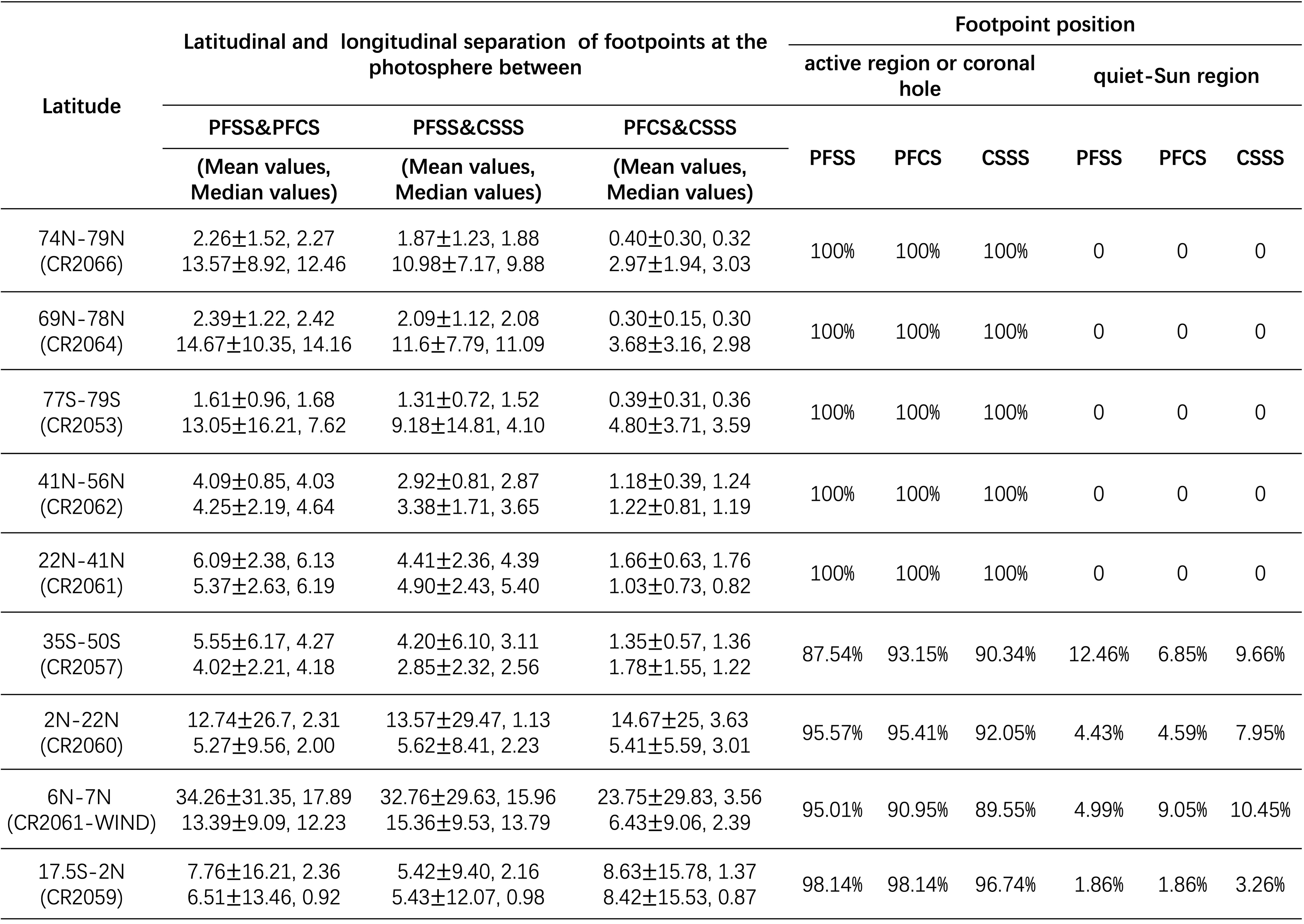}
  \label{tab:2}
\end{table}

\begin{table}[htbp]
  \centering
  \caption{Separation and Position of Photospheric Footpoints among the PFSS, PFCS, and CSSS CMF Models under the Parker HMF Model.}
  \includegraphics[width=1\textwidth]{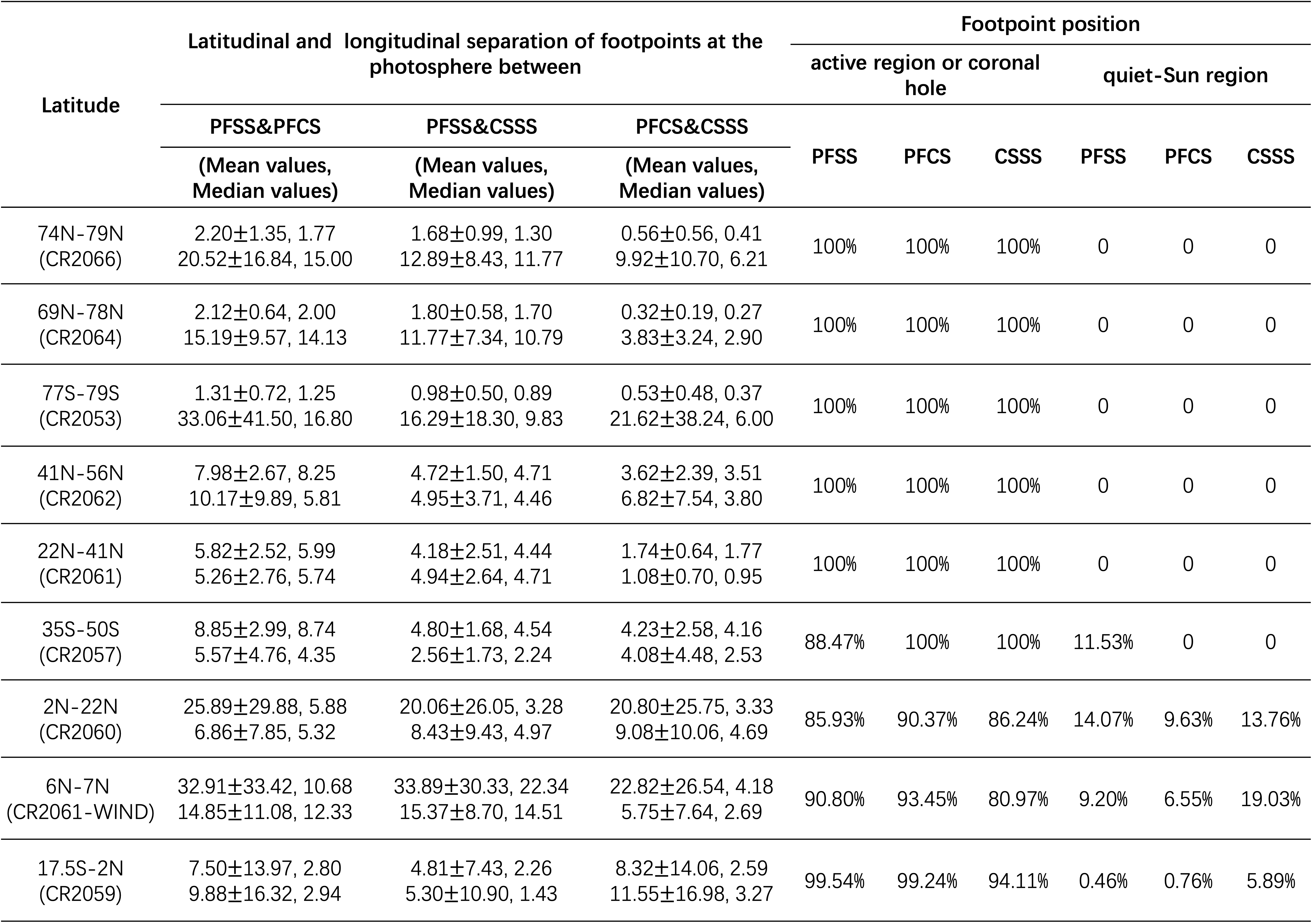}
  \label{tab:3}
\end{table}

With the HMF model held constant, the PFSS, PFCS, and CSSS CMF models yield significantly different solar wind source region derivations (see Tables \ref{tab:2} and \ref{tab:3}). The latitudinal separations ($\Delta\theta$) show a distinct pattern. At high latitudes, both the mean and median $\Delta\theta$ across models are typically on the order of several degrees. At low latitudes, the averaged $\Delta\theta$ reaches several tens of degrees, while the median $\Delta\theta$ is several degrees in most cases. Longitudinal variations ($\Delta\phi$) demonstrate another pattern: at both high and low latitudes, the mean and median $\Delta\phi$ are about tens of degrees, while at mid-latitude, both statistical measures diminish to only several degrees. Furthermore, a comparison of model agreements shows that the PFCS and CSSS models exhibit the closest agreement in both mean and median deviations, whereas the PFSS and PFCS models show the greatest divergence across all metrics.

When coupling the PFSS CMF model with the Parker HMF model, we find that the source surface height $R_\mathrm{SS}$ modulates the identification of solar wind source regions. Variations in $R_\mathrm{SS}$ have minimal impact on predicting mid-latitude HMF polarity and strength, as evidenced by both the mean and median latitudinal and longitudinal deviations between successive heights remaining within a few degrees and showing close agreement. This indicates a convergence toward consistent mid-latitude solar wind sources for $R_\mathrm{SS} \geq 2.5R_{\odot}$. In contrast, ecliptic solar wind sources exhibit a pronounced dependence on $R_\mathrm{SS}$. At $R_\mathrm{SS} = 2.0R_{\odot}$, for instance, the solar wind maps to mid-latitudes or quiet-Sun regions, whereas higher $R_\mathrm{SS}$ values map it to high latitudes or active regions. The median latitudinal deviation in the ecliptic, while smaller than the mean, remains substantially larger than the mid-latitude values. Notably, angular deviations between different $R_{\mathrm{SS}}$ pairs follow the bimodal distribution: most frequently small separations coexist with less frequent but substantial divergences ($>50^\circ$).

This study provides a comprehensive quantification of how the interplay between coronal and heliospheric magnetic field models influences the uncertainty and reliability of solar wind source region identification. The inherent model-dependent variability should be acknowledged and quantified when interpreting backmapping results for specific solar wind streams or correlating in situ measurements with solar source features. For polar coronal hole wind studies, simple models like PFSS coupled with Parker may be reasonably adequate, supported by the relative robustness of high-latitude source identification across the CMF models and the higher consistency of the HMF predictions with measurements. Conversely, low-latitude/ecliptic solar wind exhibits the heightened sensitivity to model choices, underscoring the challenges in pinpointing the origins of the slow solar wind stemming from its complex and likely multi-source nature \citep{2011Riley,2015Fu,2016Abbo,2017Cranmer,2020Viall,2021Tian}. Our future works include to investigate the formation and evolution of co-rotation interaction region and their impact on solar wind backmapping using the adaptive mesh refinement solar-interplanetary space-time conservation element and solution element (AMR-SIP-CESE) solar wind numerical model \citep{Feng2010, Feng2012, Yang2012, Yang2023, Yang2025}.

In conclusion, while models provide essential tools for connecting the Sun and the heliosphere, our results demonstrate that significant uncertainties in solar wind source identification arise directly from the choice of coronal and heliospheric magnetic field models and their parameters. Acknowledging, quantifying, and understanding these uncertainties, as initiated here, is crucial for the accurate interpretation of solar wind measurements and the advancement of reliable space weather prediction.

%% Please use the acknowledgment and contribution environments. This will
%% be anonomyized when the "anonymous" style option is used.
\begin{acknowledgments}
  
This work is supported by the National Key R$\&$D Program of China (Nos. 2022YFF0503800), China's Space Origins Exploration Program, the National Natural Science Foundation of China (Nos. 42274213, 42474216, 42030204) and the Climbing Program of NSSC (E4PD3001). The work was carried out at National Supercomputer Center in Tianjin, China, and the calculations were performed on TianHe-1 (A). We gratefully acknowledge the use of data from the following sources. We thank the Global Oscillation Network Group (GONG) for their open-data use policy. We thank the SPDF team for providing these data through the COHOWeb service. We acknowledge the Ulysses Data System (UDS) for providing the data utilized in this study. We thank the NASA Wind mission team for the Wind spacecraft data used in this study, obtained via the NASA CDAWeb service (https://cdaweb.gsfc.nasa.gov). Data were provided by the STEREO/SECCHI consortium, comprising NRL, LMSAL, NASA/GSFC (US), RAL, UBHAM (UK), MPS (Germany), CSL (Belgium), IOTA, and IAS (France).

\end{acknowledgments}


%% For this sample we use BibTeX plus aasjournalv7.bst to generate the

\bibliography{sample701}{}

\begin{thebibliography}{}
\expandafter\ifx\csname natexlab\endcsname\relax\def\natexlab#1{#1}\fi
\providecommand{\url}[1]{\href{#1}{#1}}
\providecommand{\dodoi}[1]{doi:~\href{http://doi.org/#1}{\nolinkurl{#1}}}
\providecommand{\doeprint}[1]{\href{http://ascl.net/#1}{\nolinkurl{http://ascl.net/#1}}}
\providecommand{\doarXiv}[1]{\href{https://arxiv.org/abs/#1}{\nolinkurl{https://arxiv.org/abs/#1}}}

% type= article
\bibitem[{L. {Abbo} {et~al.}(2016){Abbo}, {Ofman}, {Antiochos}, {Hansteen}, {Harra}, {Ko}, {Lapenta}, {Li}, {Riley}, {Strachan}, {von Steiger}, \& {Wang}}]{2016Abbo}
{Abbo}, L., {Ofman}, L., {Antiochos}, S.~K., {et~al.} 2016, \bibinfo{title}{{Slow Solar Wind: Observations and Modeling},} \ssr, 201, 55, \dodoi{10.1007/s11214-016-0264-1}

% type= article
\bibitem[{M. {Altschuler} \& G. {Newkirk}(1969){Altschuler} \& {Newkirk}}]{1969Altschuler}
{Altschuler}, M., \& {Newkirk}, G. 1969, \bibinfo{title}{Magnetic fields and the structure of the solar corona,} Sol Phys, 9, 131, \dodoi{10.1007/BF00145734}

% type= article
\bibitem[{C.~N. Arge {et~al.}(2010)Arge, Henney, Koller, Compeau, Young, MacKenzie, Fay, \& Harvey}]{Arge2010}
Arge, C.~N., Henney, C.~J., Koller, J., {et~al.} 2010, \bibinfo{title}{Air Force Data Assimilative Photospheric Flux Transport (ADAPT) Model,} AIP Conference Proceedings, 1216, 343, \dodoi{10.1063/1.3395870}

% type= article
\bibitem[{C.~N. Arge \& V.~J. Pizzo(2000)Arge \& Pizzo}]{2000Arge}
Arge, C.~N., \& Pizzo, V.~J. 2000, \bibinfo{title}{Improvement in the prediction of solar wind conditions using near-real time solar magnetic field updates,} Journal of Geophysical Research: Space Physics, 105, 10465, \dodoi{https://doi.org/10.1029/1999JA000262}

% type= article
\bibitem[{S.~D. {Bale} {et~al.}(2023){Bale}, {Drake}, {McManus}, {Desai}, {Badman}, {Larson}, {Swisdak}, {Horbury}, {Raouafi}, {Phan}, {Velli}, {McComas}, {Cohen}, {Mitchell}, {Panasenco}, \& {Kasper}}]{Bale2023}
{Bale}, S.~D., {Drake}, J.~F., {McManus}, M.~D., {et~al.} 2023, \bibinfo{title}{{Interchange reconnection as the source of the fast solar wind within coronal holes},} \nat, 618, 252, \dodoi{10.1038/s41586-023-05955-3}

% type= article
\bibitem[{A. {Balogh} {et~al.}(1992){Balogh}, {Beek}, {Forsyth}, {Hedgecock}, {Marquedant}, {Smith}, {Southwood}, \& {Tsurutani}}]{1992Balogh}
{Balogh}, A., {Beek}, T.~J., {Forsyth}, R.~J., {et~al.} 1992, \bibinfo{title}{{The magnetic field investigation on the ULYSSES mission - Instrumentation and preliminary scientific results},} \aaps, 92, 221

% type= article
\bibitem[{A. {Balogh} \& G. {Erd{\~o}s}(2013){Balogh} \& {Erd{\~o}s}}]{2013Balogh}
{Balogh}, A., \& {Erd{\~o}s}, G. 2013, \bibinfo{title}{{The Heliospheric Magnetic Field},} \ssr, 176, 177, \dodoi{10.1007/s11214-011-9835-3}

% type= article
\bibitem[{N. {Bizien} {et~al.}(2025){Bizien}, {Froment}, {Madjarska}, {Dudok de Wit}, \& {Velli}}]{Bizien2025}
{Bizien}, N., {Froment}, C., {Madjarska}, M.~S., {Dudok de Wit}, T., \& {Velli}, M. 2025, \bibinfo{title}{{Tracing magnetic switchbacks to their source: An assessment of solar coronal jets as switchback precursors},} \aap, 694, A181, \dodoi{10.1051/0004-6361/202452140}

% type= article
\bibitem[{R.~A. {Burger} {et~al.}(2008){Burger}, {Krüger}, {Hitge}, \& {Engelbrecht}}]{2008Burger}
{Burger}, R.~A., {Krüger}, T. P.~J., {Hitge}, M., \& {Engelbrecht}, N.~E. 2008, \bibinfo{title}{A Fisk-Parker Hybrid Heliospheric Magnetic Field with a Solar-Cycle Dependence,} The Astrophysical Journal, 674, 511, \dodoi{10.1086/525039}

% type= misc
\bibitem[{S.~R. Cranmer(2025)Cranmer}]{2025Cranmer}
Cranmer, S.~R. 2025, Solar Wind Origin, \doarXiv{2507.13460}

% type= article
\bibitem[{S.~R. {Cranmer} {et~al.}(2017){Cranmer}, {Gibson}, \& {Riley}}]{2017Cranmer}
{Cranmer}, S.~R., {Gibson}, S.~E., \& {Riley}, P. 2017, \bibinfo{title}{{Origins of the Ambient Solar Wind: Implications for Space Weather},} \ssr, 212, 1345, \dodoi{10.1007/s11214-017-0416-y}

% type= article
\bibitem[{J.-B. {Dakeyo} {et~al.}(2024){Dakeyo}, {Badman}, {Rouillard}, {Réville}, {Verscharen}, {Démoulin}, \& {Maksimovic}}]{2024Dakeyo}
{Dakeyo}, J.-B., {Badman}, S.~T., {Rouillard}, A.~P., {et~al.} 2024, \bibinfo{title}{Radial evolution of the accuracy of ballistic solar wind backmapping,} A\&A, 686, A12, \dodoi{10.1051/0004-6361/202348892}

% type= article
\bibitem[{Y. {Deng} {et~al.}(2025){Deng}, {Tian}, {Jiang}, Shuhong, Hao, Robert, Laurent, Louise, F, Frédéric, Xianyong, Luis, Linjie, Pengfei, Pradeep, Jackie, Fabio, Li, Xueshang, Weiqun, Don, Jiansen, Junfeng, Zhenyong, Chunlan, Wenya, Jiaben, Dibyendu, Vaibhav, Marco, Taro, Sayamanthula, Fang, Yang, Shin, Durgesh, Linghua, JingJing, Lidong, Ming, Yihua, Liping, Shangbin, Mei, Guiping, Xiaoshuai, Jingxiu, \& Chi}]{2025Deng}
{Deng}, Y., {Tian}, H., {Jiang}, J., {et~al.} 2025, \bibinfo{title}{Probing Solar Polar Regions,} Chinese Journal of Space Science, 45, 1, \dodoi{10.11728/cjss2025.04.2025-0054}

% type= article
\bibitem[{T. {Ervin} {et~al.}(2024){Ervin}, {Jaffarove}, {Badman}, {Huang}, {Rivera}, \& {Bale}}]{Ervin2024}
{Ervin}, T., {Jaffarove}, K., {Badman}, S.~T., {et~al.} 2024, \bibinfo{title}{{Characteristics and Source Regions of Slow Alfv{\'e}nic Solar Wind Observed by Parker Solar Probe},} \apj, 975, 156, \dodoi{10.3847/1538-4357/ad7d00}

% type= article
\bibitem[{X. {Feng} {et~al.}(2012){Feng}, {Yang}, {Xiang}, {Jiang}, {Ma}, {Wu}, {Zhong}, \& {Zhou}}]{Feng2012}
{Feng}, X., {Yang}, L., {Xiang}, C., {et~al.} 2012, \bibinfo{title}{{Validation of the 3D AMR SIP-CESE Solar Wind Model for Four Carrington Rotations},} \solphys, 279, 207, \dodoi{10.1007/s11207-012-9969-9}

% type= article
\bibitem[{X. {Feng} {et~al.}(2010){Feng}, {Yang}, {Xiang}, {Wu}, {Zhou}, \& {Zhong}}]{Feng2010}
{Feng}, X., {Yang}, L., {Xiang}, C., {et~al.} 2010, \bibinfo{title}{{Three-dimensional Solar WIND Modeling from the Sun to Earth by a SIP-CESE MHD Model with a Six-component Grid},} \apj, 723, 300, \dodoi{10.1088/0004-637X/723/1/300}

% type= article
\bibitem[{L.~A. {Fisk}(1996){Fisk}}]{1996Fisk}
{Fisk}, L.~A. 1996, \bibinfo{title}{Motion of the footpoints of heliospheric magnetic field lines at the Sun: Implications for recurrent energetic particle events at high heliographic latitudes,} Journal of Geophysical Research: Space Physics, 101, 15547, \dodoi{https://doi.org/10.1029/96JA01005}

% type= article
\bibitem[{L.~A. Fisk {et~al.}(1999)Fisk, Zurbuchen, \& Schwadron}]{1999Fisk}
Fisk, L.~A., Zurbuchen, T.~H., \& Schwadron, N.~A. 1999, \bibinfo{title}{On the Coronal Magnetic Field: Consequences of Large-Scale Motions,} The Astrophysical Journal, 521, 868, \dodoi{10.1086/307556}

% type= article
\bibitem[{R.~J. {Forsyth} {et~al.}(2002){Forsyth}, {Balogh}, \& {Smith}}]{2002Forsyth}
{Forsyth}, R.~J., {Balogh}, A., \& {Smith}, E.~J. 2002, \bibinfo{title}{{The underlying direction of the heliospheric magnetic field through the Ulysses first orbit},} Journal of Geophysical Research (Space Physics), 107, 1405, \dodoi{10.1029/2001JA005056}

% type= article
\bibitem[{H. {Fu} {et~al.}(2015){Fu}, {Li}, {Li}, {Huang}, {Mou}, Jiao, \& Xia}]{2015Fu}
{Fu}, H., {Li}, B., {Li}, X., {et~al.} 2015, \bibinfo{title}{Coronal Sources and In Situ Properties of the Solar Winds Sampled by ACE During 1999-2008,} Solar Physics, 290, 1399, \dodoi{10.1007/s11207-015-0689-9}

% type= article
\bibitem[{C. {Hou} {et~al.}(2024{\natexlab{a}}){Hou}, {He}, {Duan}, {Wu}, {Chen}, {Verscharen}, {Rouillard}, {Li}, {Yang}, \& {Bale}}]{Hou2024a}
{Hou}, C., {He}, J., {Duan}, D., {et~al.} 2024{\natexlab{a}}, \bibinfo{title}{{The origin of interplanetary switchbacks in reconnection at chromospheric network boundaries},} Nature Astronomy, 8, 1246, \dodoi{10.1038/s41550-024-02321-9}

% type= article
\bibitem[{C. {Hou} {et~al.}(2024{\natexlab{b}}){Hou}, {Rouillard}, {He}, {Gannouni}, {R{\'e}ville}, {Louarn}, {Fedorov}, {P{\v{r}}ech}, {Owen}, {Verscharen}, {D'Amicis}, {Sorriso-Valvo}, {Fargette}, {Coburn}, {G{\'e}not}, {Raines}, {Bruno}, {Livi}, {Lavraud}, {Andr{\'e}}, {Fruit}, {Kieokaew}, {Plotnikov}, {Penou}, {Barthe}, {Kataria}, {Berthomier}, {Allegrini}, {Fortunato}, {Mele}, \& {Horbury}}]{Hou2024b}
{Hou}, C., {Rouillard}, A.~P., {He}, J., {et~al.} 2024{\natexlab{b}}, \bibinfo{title}{{Connecting Solar Wind Velocity Spikes Measured by Solar Orbiter and Coronal Brightenings Observed by SDO},} \apjl, 968, L28, \dodoi{10.3847/2041-8213/ad4eda}

% type= article
\bibitem[{R.~A. {Howard} {et~al.}(2008){Howard}, {Moses}, {Vourlidas}, {Newmark}, {Socker}, {Plunkett}, {Korendyke}, {Cook}, {Hurley}, {Davila}, {Thompson}, {St Cyr}, {Mentzell}, {Mehalick}, {Lemen}, {Wuelser}, {Duncan}, {Tarbell}, {Wolfson}, {Moore}, {Harrison}, {Waltham}, {Lang}, {Davis}, {Eyles}, {Mapson-Menard}, {Simnett}, {Halain}, {Defise}, {Mazy}, {Rochus}, {Mercier}, {Ravet}, {Delmotte}, {Auchere}, {Delaboudiniere}, {Bothmer}, {Deutsch}, {Wang}, {Rich}, {Cooper}, {Stephens}, {Maahs}, {Baugh}, {McMullin}, \& {Carter}}]{2008Howard}
{Howard}, R.~A., {Moses}, J.~D., {Vourlidas}, A., {et~al.} 2008, \bibinfo{title}{Sun Earth Connection Coronal and Heliospheric Investigation (SECCHI),} Space Science Reviews, 136, 67, \dodoi{10.1007/s11214-008-9341-4}

% type= book
\bibitem[{J.~D. {Jackson}(1962){Jackson}}]{1962Jackson}
{Jackson}, J.~D. 1962, Classical Electrodynamics (Wiley).
\newblock \url{https://ui.adsabs.harvard.edu/abs/1962clel.book.....J}

% type= misc
\bibitem[{J.~H. {King} \& N.~E. {Papitashvili}(2020){King} \& {Papitashvili}}]{King_Papitashvili_2020}
{King}, J.~H., \& {Papitashvili}, N.~E. 2020, OMNI Combined Heliopheric Observations (COHO), Merged Magnetic Field, Plasma and Ephermeris, Definitive Hourly Data, \dodoi{10.48322/6FFX-3441}

% type= article
\bibitem[{J. {Koskela} {et~al.}(2019){Koskela}, {Virtanen}, \& {Mursula}}]{2019Koskela}
{Koskela}, J., {Virtanen}, I., \& {Mursula}, K. 2019, \bibinfo{title}{{Revisiting the coronal current sheet model: Parameter range analysis and comparison with the potential field model},} \aap, 631, A17, \dodoi{10.1051/0004-6361/201935967}

% type= article
\bibitem[{A. Koukras {et~al.}(2025)Koukras, Dolla, \& Keppens}]{2025Koukras}
Koukras, A., Dolla, L., \& Keppens, R. 2025, \bibinfo{title}{Estimating uncertainties in the back-mapping of the fast solar wind,} Astronomy \& Astrophysics, 694, A134, \dodoi{10.1051/0004-6361/202244327}

% type= article
\bibitem[{P. {Kumar} {et~al.}(2023){Kumar}, {Karpen}, {Uritsky}, {Deforest}, {Raouafi}, {DeVore}, \& {Antiochos}}]{Kumar2023}
{Kumar}, P., {Karpen}, J.~T., {Uritsky}, V.~M., {et~al.} 2023, \bibinfo{title}{{New Evidence on the Origin of Solar Wind Microstreams/Switchbacks},} \apjl, 951, L15, \dodoi{10.3847/2041-8213/acd54e}

% type= article
\bibitem[{R.~H. {Levine} {et~al.}(1977){Levine}, {Altschuler}, \& {Harvey}}]{Levine1977}
{Levine}, R.~H., {Altschuler}, M.~D., \& {Harvey}, J.~W. 1977, \bibinfo{title}{{Solar sources of the interplanetary magnetic field and solar wind},} \jgr, 82, 1061, \dodoi{10.1029/JA082i007p01061}

% type= article
\bibitem[{H. {Li} {et~al.}(2021){Li}, {Feng}, \& {Wei}}]{Li2021}
{Li}, H., {Feng}, X., \& {Wei}, F. 2021, \bibinfo{title}{{Comparison of Synoptic Maps and PFSS Solutions for The Declining Phase of Solar Cycle 24},} Journal of Geophysical Research (Space Physics), 126, e28870, \dodoi{10.1029/2020JA028870}

% type= article
\bibitem[{R. Lin {et~al.}(2024)Lin, Luo, He, Xie, Hou, \& Chen}]{2023Lin}
Lin, R., Luo, Z., He, J., {et~al.} 2024, \bibinfo{title}{Prediction of Solar Wind Speed Through Machine Learning From Extrapolated Solar Coronal Magnetic Field,} Space Weather, 22, e2023SW003561, \dodoi{https://doi.org/10.1029/2023SW003561}

% type= article
\bibitem[{J.~A. {Linker} {et~al.}(1999){Linker}, {Miki{\'c}}, {Biesecker}, {Forsyth}, {Gibson}, {Lazarus}, {Lecinski}, {Riley}, {Szabo}, \& {Thompson}}]{Linker1999}
{Linker}, J.~A., {Miki{\'c}}, Z., {Biesecker}, D.~A., {et~al.} 1999, \bibinfo{title}{{Magnetohydrodynamic modeling of the solar corona during Whole Sun Month},} \jgr, 104, 9809, \dodoi{10.1029/1998JA900159}

% type= article
\bibitem[{R. Lionello {et~al.}(2005)Lionello, Riley, Linker, \& Mikić}]{2005Lionello}
Lionello, R., Riley, P., Linker, J.~A., \& Mikić, Z. 2005, \bibinfo{title}{The Effects of Differential Rotation on the Magnetic Structure of the Solar Corona: Magnetohydrodynamic Simulations,} The Astrophysical Journal, 625, 463, \dodoi{10.1086/429268}

% type= article
\bibitem[{A.~R. Macneil {et~al.}(2021)Macneil, Owens, Finley, \& Matt}]{2021Macneil}
Macneil, A.~R., Owens, M.~J., Finley, A.~J., \& Matt, S.~P. 2021, \bibinfo{title}{A statistical evaluation of ballistic backmapping for the slow solar wind: the interplay of solar wind acceleration and corotation,} Monthly Notices of the Royal Astronomical Society, 509, 2390, \dodoi{10.1093/mnras/stab2965}

% type= article
\bibitem[{D. {M{\"u}ller} {et~al.}(2020){M{\"u}ller}, {St. Cyr}, {Zouganelis}, {Gilbert}, {Marsden}, {Nieves-Chinchilla}, {Antonucci}, {Auch{\`e}re}, {Berghmans}, {Horbury}, {Howard}, {Krucker}, {Maksimovic}, {Owen}, {Rochus}, {Rodriguez-Pacheco}, {Romoli}, {Solanki}, {Bruno}, {Carlsson}, {Fludra}, {Harra}, {Hassler}, {Livi}, {Louarn}, {Peter}, {Sch{\"u}hle}, {Teriaca}, {del Toro Iniesta}, {Wimmer-Schweingruber}, {Marsch}, {Velli}, {De Groof}, {Walsh}, \& {Williams}}]{2020Muller}
{M{\"u}ller}, D., {St. Cyr}, O.~C., {Zouganelis}, I., {et~al.} 2020, \bibinfo{title}{{The Solar Orbiter mission. Science overview},} \aap, 642, A1, \dodoi{10.1051/0004-6361/202038467}

% type= article
\bibitem[{M. {Neugebauer} {et~al.}(1998){Neugebauer}, {Forsyth}, {Galvin}, {Harvey}, {Hoeksema}, {Lazarus}, {Lepping}, {Linker}, {Mikic}, {Steinberg}, {von Steiger}, {Wang}, \& {Wimmer-Schweingruber}}]{Neugebauer1998}
{Neugebauer}, M., {Forsyth}, R.~J., {Galvin}, A.~B., {et~al.} 1998, \bibinfo{title}{{Spatial structure of the solar wind and comparisons with solar data and models},} \jgr, 103, 14587, \dodoi{10.1029/98JA00798}

% type= article
\bibitem[{M. Neugebauer {et~al.}(1998)Neugebauer, Forsyth, Galvin, Harvey, Hoeksema, Lazarus, Lepping, Linker, Mikic, Steinberg, von Steiger, Wang, \& Wimmer-Schweingruber}]{1998Neugebauer}
Neugebauer, M., Forsyth, R.~J., Galvin, A.~B., {et~al.} 1998, \bibinfo{title}{Spatial structure of the solar wind and comparisons with solar data and models,} Journal of Geophysical Research: Space Physics, 103, 14587, \dodoi{https://doi.org/10.1029/98JA00798}

% type= article
\bibitem[{J. Nolte \& E. Roelof(1973)Nolte \& Roelof}]{1973Nolte}
Nolte, J., \& Roelof, E. 1973, \bibinfo{title}{Large-scale structure of the interplanetary medium.,} Sol Phys, 33, 241, \dodoi{10.1007/BF00152395}

% type= article
\bibitem[{M. Owens \& R. Forsyth(2013)Owens \& Forsyth}]{2013Owens}
Owens, M., \& Forsyth, R. 2013, \bibinfo{title}{The Heliospheric Magnetic Field.,} Sol. Phys., 10, \dodoi{https://doi.org/10.12942/lrsp-2013-5}

% type= article
\bibitem[{S. {Parenti} {et~al.}(2021){Parenti}, {Chifu}, {Del Zanna}, {Edmondson}, {Giunta}, {Hansteen}, {Higginson}, {Laming}, {Lepri}, {Lynch}, {Rivera}, {von Steiger}, {Wiegelmann}, {Wimmer-Schweingruber}, {Zambrana Prado}, \& {Pelouze}}]{Parenti2021}
{Parenti}, S., {Chifu}, I., {Del Zanna}, G., {et~al.} 2021, \bibinfo{title}{{Linking the Sun to the Heliosphere Using Composition Data and Modelling},} \ssr, 217, 78, \dodoi{10.1007/s11214-021-00856-1}

% type= article
\bibitem[{J.-S. Park {et~al.}(2025)Park, Shi, Seough, Zhang, Tang, Guo, Cho, Lee, \& Kim}]{Park2025}
Park, J.-S., Shi, Q., Seough, J., {et~al.} 2025, \bibinfo{title}{Non-Parker Spiral Interplanetary Magnetic Field Configurations Observed in Near-Earth Space: Statistical Features of Their Occurrence and Solar Wind Conditions,} The Astrophysical Journal Supplement Series, 280, 2, \dodoi{10.3847/1538-4365/adea79}

% type= article
\bibitem[{E.~N. {Parker}(1958){Parker}}]{1958Parker}
{Parker}, E.~N. 1958, \bibinfo{title}{Dynamics of the Interplanetary Gas and Magnetic Fields,} \apj, 128, 664, \dodoi{10.1086/146579}

% type= article
\bibitem[{T. Peleikis {et~al.}(2017)Peleikis, Kruse, Berger, \& Wimmer–Schweingruber}]{2017Peleikis}
Peleikis, T., Kruse, M., Berger, L., \& Wimmer–Schweingruber, R.~F. 2017, \bibinfo{title}{Origin of the solar wind: A novel approach to link in situ and remote observations - A study for SPICE and SWA on the upcoming Solar Orbiter mission,} Astronomy and Astrophysics, 602.
\newblock \url{https://api.semanticscholar.org/CorpusID:59431158}

% type= article
\bibitem[{J. Phillips {et~al.}(1995)Phillips, Bame, Feldman, Gosling, Hammond, McComas, Goldstein, \& Neugebauer}]{1995PHILLIPS}
Phillips, J., Bame, S., Feldman, W., {et~al.} 1995, \bibinfo{title}{Ulysses solar wind plasma observations during the declining phase of solar cycle 22,} Advances in Space Research, 16, 85, \dodoi{https://doi.org/10.1016/0273-1177(95)00318-9}

% type= article
\bibitem[{V.~J. {Pizzo}(1981){Pizzo}}]{Pizzo1981}
{Pizzo}, V.~J. 1981, \bibinfo{title}{{On the application of numerical models to the inverse mapping of solar wind flow structures},} \jgr, 86, 6685, \dodoi{10.1029/JA086iA08p06685}

% type= article
\bibitem[{A. {Posner} {et~al.}(2001){Posner}, {Zurbuchen}, {Schwadron}, {Fisk}, {Gloeckler}, {Linker}, {Miki{\'c}}, \& {Riley}}]{Posner2001}
{Posner}, A., {Zurbuchen}, T.~H., {Schwadron}, N.~A., {et~al.} 2001, \bibinfo{title}{{Nature of the boundary between open and closed magnetic field line regions at the Sun revealed by composition data and numerical models},} \jgr, 106, 15869, \dodoi{10.1029/2000JA000112}

% type= article
\bibitem[{M.~A. Reiss {et~al.}(2021)Reiss, Muglach, Möstl, Arge, Bailey, Delouille, Garton, Hamada, Hofmeister, Illarionov, Jarolim, Kirk, Kosovichev, Krista, Lee, Lowder, MacNeice, Veronig, \& Team}]{2021Reiss}
Reiss, M.~A., Muglach, K., Möstl, C., {et~al.} 2021, \bibinfo{title}{The Observational Uncertainty of Coronal Hole Boundaries in Automated Detection Schemes,} The Astrophysical Journal, 913, 28, \dodoi{10.3847/1538-4357/abf2c8}

% type= article
\bibitem[{P. {Riley} \& R. {Lionello}(2011){Riley} \& {Lionello}}]{2011Riley}
{Riley}, P., \& {Lionello}, R. 2011, \bibinfo{title}{Mapping Solar Wind Streams from the Sun to 1 AU: A Comparison of Techniques,} Solar Physics, 270, 575, \dodoi{10.1007/s11207-011-9766-x}

% type= article
\bibitem[{Y.~J. {Rivera} {et~al.}(2024){Rivera}, {Badman}, {Stevens}, {Verniero}, {Stawarz}, {Shi}, {Raines}, {Paulson}, {Owen}, {Niembro}, {Louarn}, {Livi}, {Lepri}, {Kasper}, {Horbury}, {Halekas}, {Dewey}, {De Marco}, \& {Bale}}]{Rivera2024}
{Rivera}, Y.~J., {Badman}, S.~T., {Stevens}, M.~L., {et~al.} 2024, \bibinfo{title}{{In situ observations of large-amplitude Alfv{\'e}n waves heating and accelerating the solar wind},} Science, 385, 962, \dodoi{10.1126/science.adk6953}

% type= article
\bibitem[{A.~P. {Rouillard} {et~al.}(2020){Rouillard}, {Pinto}, {Vourlidas}, {De Groof}, {Thompson}, {Bemporad}, {Dolei}, {Indurain}, {Buchlin}, {Sasso}, {Spadaro}, {Dalmasse}, {Hirzberger}, {Zouganelis}, {Strugarek}, {Brun}, {Alexandre}, {Berghmans}, {Raouafi}, {Wiegelmann}, {Pagano}, {Arge}, {Nieves-Chinchilla}, {Lavarra}, {Poirier}, {Amari}, {Aran}, {Andretta}, {Antonucci}, {Anastasiadis}, {Auch{\`e}re}, {Bellot Rubio}, {Nicula}, {Bonnin}, {Bouchemit}, {Budnik}, {Caminade}, {Cecconi}, {Carlyle}, {Cernuda}, {Davila}, {Etesi}, {Espinosa Lara}, {Fedorov}, {Fineschi}, {Fludra}, {G{\'e}not}, {Georgoulis}, {Gilbert}, {Giunta}, {Gomez-Herrero}, {Guest}, {Haberreiter}, {Hassler}, {Henney}, {Howard}, {Horbury}, {Janvier}, {Jones}, {Kozarev}, {Kraaikamp}, {Kouloumvakos}, {Krucker}, {Lagg}, {Linker}, {Lavraud}, {Louarn}, {Maksimovic}, {Maloney}, {Mann}, {Masson}, {M{\"u}ller}, {{\"O}nel}, {Osuna}, {Orozco Suarez}, {Owen}, {Papaioannou}, {P{\'e}rez-Su{\'a}rez}, {Rodriguez-Pacheco}, {Parenti}, {Pariat}, {Peter}, {Plunkett}, {Pomoell}, {Raines}, {Riethm{\"u}ller}, {Rich}, {Rodriguez}, {Romoli}, {Sanchez}, {Solanki}, {St Cyr}, {Straus}, {Susino}, {Teriaca}, {del Toro Iniesta}, {Ventura}, {Verbeeck}, {Vilmer}, {Warmuth}, {Walsh}, {Watson}, {Williams}, {Wu}, \& {Zhukov}}]{Rouillard2020}
{Rouillard}, A.~P., {Pinto}, R.~F., {Vourlidas}, A., {et~al.} 2020, \bibinfo{title}{{Models and data analysis tools for the Solar Orbiter mission},} \aap, 642, A2, \dodoi{10.1051/0004-6361/201935305}

% type= article
\bibitem[{K.~H. {Schatten}(1971){Schatten}}]{1971Schatten}
{Schatten}, K.~H. 1971, \bibinfo{title}{Current sheet magnetic model for the solar corona,} Cosmic Electrodynamics, 2, 232

% type= article
\bibitem[{K.~H. {Schatten} {et~al.}(1969){Schatten}, {Wilcox}, \& {Ness}}]{1969Schatten}
{Schatten}, K.~H., {Wilcox}, J.~M., \& {Ness}, N.~F. 1969, \bibinfo{title}{A model of interplanetary and coronal magnetic fields,} Solar Physics, 6, 442, \dodoi{10.1007/BF00146478}

% type= article
\bibitem[{N.~A. {Schwadron} \& D.~J. {McComas}(2003){Schwadron} \& {McComas}}]{2003Schwadron}
{Schwadron}, N.~A., \& {McComas}, D.~J. 2003, \bibinfo{title}{Heliospheric “FALTS�? Favored Acceleration Locations at the Termination Shock,} Geophysical Research Letters, 30, \dodoi{https://doi.org/10.1029/2002GL016499}

% type= article
\bibitem[{G. Shi {et~al.}(2024)Shi, Feng, Ying, Li, \& Gan}]{2024Shi}
Shi, G., Feng, L., Ying, B., Li, S., \& Gan, W. 2024, \bibinfo{title}{Refinement of Global Coronal and Interplanetary Magnetic Field Extrapolations Constrained by Remote-sensing and In Situ Observations at the Solar Minimum,} The Astrophysical Journal, 970, 131, \dodoi{10.3847/1538-4357/ad5200}

% type= article
\bibitem[{X. {Shi} {et~al.}(2024){Shi}, {Fu}, {Huang}, {Yan}, {Ma}, {Huangfu}, {Song}, \& {Xia}}]{2024ShiXinzheng}
{Shi}, X., {Fu}, H., {Huang}, Z., {et~al.} 2024, \bibinfo{title}{{The Differences in the Origination and Properties of the Near-Earth Solar Wind between Solar Cycles 23 and 24},} \apj, 972, 54, \dodoi{10.3847/1538-4357/ad5be1}

% type= article
\bibitem[{E.~J. {Smith} \& A. {Balogh}(1995){Smith} \& {Balogh}}]{Smith1995}
{Smith}, E.~J., \& {Balogh}, A. 1995, \bibinfo{title}{{Ulysses observations of the radial magnetic field},} \grl, 22, 3317, \dodoi{10.1029/95GL02826}

% type= article
\bibitem[{P.~J. {Steyn} \& R.~A. {Burger}(2020){Steyn} \& {Burger}}]{2020Steyn}
{Steyn}, P.~J., \& {Burger}, R.~A. 2020, \bibinfo{title}{A Generalized Fisk-type HMF: Implications of Spatially Dependent Photospheric Differential Rotation,} The Astrophysical Journal, 902, \dodoi{10.3847/1538-4357/abb2a5}

% type= article
\bibitem[{P.~J. {Steyn} {et~al.}(2024){Steyn}, {Johnson}, {Botha}, \& {R{\'e}gnier}}]{2024Steyn}
{Steyn}, P.~J., {Johnson}, D., {Botha}, G.~J.~J., \& {R{\'e}gnier}, S. 2024, \bibinfo{title}{Identifying Coronal Sources of L1 Solar Wind Disturbances Using the Fisk Heliospheric Magnetic Field and Potential Field Extrapolations during Three Solar Minima,} \apj, 966, 77, \dodoi{10.3847/1538-4357/ad3356}

% type= article
\bibitem[{H. {Tian} {et~al.}(2021){Tian}, {Harra}, {Baker}, {Brooks}, \& {Xia}}]{2021Tian}
{Tian}, H., {Harra}, L., {Baker}, D., {Brooks}, D.~H., \& {Xia}, L. 2021, \bibinfo{title}{{Upflows in the Upper Solar Atmosphere},} \solphys, 296, 47, \dodoi{10.1007/s11207-021-01792-7}

% type= article
\bibitem[{N.~M. {Viall} \& J.~E. {Borovsky}(2020){Viall} \& {Borovsky}}]{2020Viall}
{Viall}, N.~M., \& {Borovsky}, J.~E. 2020, \bibinfo{title}{{Nine Outstanding Questions of Solar Wind Physics},} Journal of Geophysical Research (Space Physics), 125, e26005, \dodoi{10.1029/2018JA02600510.1002/essoar.10502606.1}

% type= article
\bibitem[{Y.~M. {Wang} \& N.~R. {Sheeley}(1992){Wang} \& {Sheeley}}]{1992Wang}
{Wang}, Y.~M., \& {Sheeley}, Jr., N.~R. 1992, \bibinfo{title}{{On Potential Field Models of the Solar Corona},} \apj, 392, 310, \dodoi{10.1086/171430}

% type= article
\bibitem[{L.~B. {Wilson} {et~al.}(2021){Wilson}, {Brosius}, {Gopalswamy}, {Nieves-Chinchilla}, {Szabo}, {Hurley}, {Phan}, {Kasper}, {Lugaz}, {Richardson}, {Chen}, {Verscharen}, {Wicks}, \& {TenBarge}}]{2021Wilson}
{Wilson}, III, L.~B., {Brosius}, A.~L., {Gopalswamy}, N., {et~al.} 2021, \bibinfo{title}{{A Quarter Century of Wind Spacecraft Discoveries},} Reviews of Geophysics, 59, e2020RG000714, \dodoi{10.1029/2020RG00071410.1002/essoar.10504309.2}

% type= article
\bibitem[{L. {Yang} {et~al.}(2023){Yang}, {Hou}, {Feng}, {He}, {Xiong}, {Zhang}, {Zhou}, {Shen}, {Zhao}, {Li}, {Yang}, \& {Liu}}]{Yang2023}
{Yang}, L., {Hou}, C., {Feng}, X., {et~al.} 2023, \bibinfo{title}{{Global Morphology Distortion of the 2021 October 9 Coronal Mass Ejection from an Ellipsoid to a Concave Shape},} \apj, 942, 65, \dodoi{10.3847/1538-4357/aca52d}

% type= article
\bibitem[{L. {Yang} {et~al.}(2025){Yang}, {Feng}, {Shen}, {Xiong}, {Shen}, {Chi}, {Wang}, {Yan}, {Ma}, {Zhou}, {Zhang}, \& {Zhao}}]{Yang2025}
{Yang}, L., {Feng}, X., {Shen}, F., {et~al.} 2025, \bibinfo{title}{{Expansion-induced Three-part Morphology of the 2021 December 4 Coronal Mass Ejection},} \apj, 981, 109, \dodoi{10.3847/1538-4357/adb313}

% type= article
\bibitem[{L.~P. {Yang} {et~al.}(2012){Yang}, {Feng}, {Xiang}, {Liu}, {Zhao}, \& {Wu}}]{Yang2012}
{Yang}, L.~P., {Feng}, X.~S., {Xiang}, C.~Q., {et~al.} 2012, \bibinfo{title}{{Time-dependent MHD modeling of the global solar corona for year 2007: Driven by daily-updated magnetic field synoptic data},} \jgr, 117, A08110, \dodoi{10.1029/2011JA017494}

% type= article
\bibitem[{Z. Yang {et~al.}(2024)Yang, Tian, Tomczyk, Liu, Gibson, Morton, \& Downs}]{2024Yangc}
Yang, Z., Tian, H., Tomczyk, S., {et~al.} 2024, \bibinfo{title}{Observing the evolution of the Sun’s global coronal magnetic field over 8 months,} Science, 386, 76, \dodoi{10.1126/science.ado2993}

% type= article
\bibitem[{Z. Yang {et~al.}(2020{\natexlab{a}})Yang, Tian, Tomczyk, Morton, Bai, Samanta, \& Chen}]{2020Yangb}
Yang, Z., Tian, H., Tomczyk, S., {et~al.} 2020{\natexlab{a}}, \bibinfo{title}{Mapping the magnetic field in the solar corona through magnetoseismology,} Science China Technological Sciences, 63, 2357, \dodoi{https://doi.org/10.1007/s11431-020-1706-9}

% type= article
\bibitem[{Z. Yang {et~al.}(2020{\natexlab{b}})Yang, Bethge, Tian, Tomczyk, Morton, Zanna, McIntosh, Karak, Gibson, Samanta, He, Chen, \& Wang}]{2020Yanga}
Yang, Z., Bethge, C., Tian, H., {et~al.} 2020{\natexlab{b}}, \bibinfo{title}{Global maps of the magnetic field in the solar corona,} Science, 369, 694, \dodoi{10.1126/science.abb4462}

% type= article
\bibitem[{X. Zhao \& J.~T. Hoeksema(1994)Zhao \& Hoeksema}]{1994Zhao}
Zhao, X., \& Hoeksema, J.~T. 1994, \bibinfo{title}{A coronal magnetic field model with horizontal volume and sheet currents,} Solar Physics, 151, 91, \dodoi{10.1007/BF00654084}

% type= article
\bibitem[{X. {Zhao} \& J.~T. {Hoeksema}(1995){Zhao} \& {Hoeksema}}]{1995Zhao}
{Zhao}, X., \& {Hoeksema}, J.~T. 1995, \bibinfo{title}{Prediction of the interplanetary magnetic field strength,} Journal of Geophysical Research: Space Physics, 100, 19, \dodoi{10.1029/94JA02266}

% type= article
\bibitem[{J.~B. Zirker(1977)Zirker}]{1977Zirker}
Zirker, J.~B. 1977, \bibinfo{title}{Coronal holes and high-speed wind streams,} Reviews of Geophysics, 15, 257, \dodoi{https://doi.org/10.1029/RG015i003p00257}

% type= article
\bibitem[{T.~H. {Zurbuchen} {et~al.}(1997){Zurbuchen}, {Schwadron}, \& {Fisk}}]{1997Zurbuchen}
{Zurbuchen}, T.~H., {Schwadron}, N.~A., \& {Fisk}, L.~A. 1997, \bibinfo{title}{Direct observational evidence for a heliospheric magnetic field with large excursions in latitude,} Journal of Geophysical Research: Space Physics, 102, 24175, \dodoi{https://doi.org/10.1029/97JA02194}

\end{thebibliography}
\bibliographystyle{aasjournalv7}



\end{document}